\documentclass[aps,prd,reprint,superscriptaddress,nofootinbib,amsfonts,amssymb,amsmath]{revtex4-1}

\usepackage{graphicx}
\usepackage{float}
\usepackage{bbold}
\usepackage{xcolor,rotating}
\usepackage{bbm}
\usepackage{dsfont}
\usepackage[colorlinks]{hyperref}

\DeclareMathAlphabet{\EuRoman}{U}{eur}{m}{n}
\SetMathAlphabet{\EuRoman}{bold}{U}{eur}{b}{n}

\allowdisplaybreaks

 \graphicspath{{./figures/}}

\newcommand{\Tr}{{\text{Tr}}}
\newcommand{\tr}{{\text{tr}}}
\def\di{\displaystyle}

\def\bg{\begin{eqnarray}\begin{array}{rcl}\displaystyle}
\def\eg{\end{array} &\di    &\di   \end{eqnarray}}
\def\bm#1{\begin{eqnarray}\begin{array}{#1}\di}
\def\bmo#1{\begin{eqnarray*}\begin{array}{#1}\di}
\def\bml#1#2{\begin{eqnarray}\begin{array}{#1}\label{#2}\di}
\def\bgo{\begin{eqnarray*}\begin{array}{rcl}\displaystyle}
\def\ego{\end{array} &\di    &\di \nonumber  \end{eqnarray*}}

\def\btensor#1#2{\renew\left#1\begin{array}{#2}\di}
\def\brtensor#1#2#3{\ren#3\left#1\begin{array}{#2}}
\def\botensor#1#2{\renew\left#1\begin{array}{#2}}
\def\etensor#1{\end{array}\right#1}

\hypersetup{
colorlinks=true,
citecolor=blue,
citebordercolor=red,
linktoc=all,
linkcolor=blue
}

\def\det{{\rm det}}

\def\id{1\!\mbox{l}}

\def\s0#1#2{\mbox{\small{$ \frac{#1}{#2} $}}}
\def\0#1#2{\frac{#1}{#2}}

\def\CO{{\mathcal O}}

\def\CT{{\mathcal T}}

\def\s{\sigma}

\def\ren#1{\renewcommand{\arraystretch}{#1}}

\def\renew{\renewcommand{\arraystretch}{1}}

\newcommand{\qqquad}{\qquad \qquad}

\begin{document}

\title{One force to rule them all: asymptotic safety of gravity with matter}

\author{Nicolai Christiansen}
\affiliation{Institut f\"ur Theoretische Physik, Universit\"at Heidelberg,
Philosophenweg 16, 69120 Heidelberg, Germany}
\author{Daniel F. Litim}
\affiliation{Department of Physics and Astronomy, University of Sussex,
Brighton, BN1 9QH, U.\,K.}       
\author{Jan M. Pawlowski}
\affiliation{Institut f\"ur Theoretische Physik, Universit\"at Heidelberg,
Philosophenweg 16, 69120 Heidelberg, Germany}
\affiliation{ExtreMe Matter Institute EMMI, GSI Helmholtzzentrum f\"ur
Schwerionenforschung mbH, Planckstr.\ 1, 64291 Darmstadt, Germany}
\author{Manuel Reichert}
\affiliation{Institut f\"ur Theoretische Physik, Universit\"at Heidelberg,
Philosophenweg 16, 69120 Heidelberg, Germany}

\begin{abstract}
  We study the asymptotic safety conjecture for quantum gravity in the
  presence of matter fields.  A general line of reasoning is put
  forward explaining why gravitons dominate the high-energy behaviour,
  largely independently of the matter fields as long as these remain
  sufficiently weakly coupled. Our considerations are put to work for
  gravity coupled to Yang-Mills theories with the help of the
  functional renormalisation group. In an expansion about flat
  backgrounds, explicit results for beta functions, fixed points,
  universal exponents, and scaling solutions are given in systematic
  approximations exploiting running propagators, vertices, and
  background couplings.  Invariably, we find that the gauge coupling
  becomes asymptotically free while the gravitational sector becomes
  asymptotically safe. The dependence on matter field multiplicities
  is weak.  We also explain how the scheme dependence, which is more
  pronounced, can be handled without changing the physics. Our
  findings offer a new interpretation of many earlier results, which
  is explained in detail.  The results generalise to theories with
  minimally coupled scalar and fermionic matter.  Some implications
  for the ultraviolet closure of the Standard Model or its extensions
  are given.
\end{abstract}

\maketitle

\section{Introduction}
\label{sec:intro}
The Standard Model of particle physics combines three of the four fundamentally known forces of Nature. 
It remains an open challenge to understand whether a quantum theory for gravity can be established 
under the same set of basic principles. 
Steven Weinberg's seminal asymptotic safety conjecture stipulates that it can,
provided the high energy behaviour of gravity is controlled by an interacting fixed point \cite{Weinberg:1980gg,Reuter:1996cp}.
By now, the scenario has become a viable contender with many applications 
ranging from particle physics to cosmology 
\cite{Niedermaier:2006ns,Litim:2006dx,Percacci:2007sz,Litim:2008tt,Litim:2011cp,Reuter:2012id}.

Fixed points for quantum gravity
have been obtained from the renormalisation group (RG) 
in increasingly sophisticated approximations 
ranging from the Einstein-Hilbert theory
\cite{Souma:1999at,Falkenberg:1996bq,Reuter:2001ag,Lauscher:2001ya,Litim:2003vp,
  Bonanno:2004sy,Fischer:2006fz,Eichhorn:2009ah,Manrique:2009uh,
  Eichhorn:2010tb,Groh:2010ta,Manrique:2010am,Manrique:2011jc,Litim:2012vz,
  Donkin:2012ud,Christiansen:2012rx,Codello:2013fpa,
  Christiansen:2014raa,Becker:2014qya,Falls:2014zba,
  Falls:2015qga,Falls:2015cta,Christiansen:2015rva,
  Gies:2015tca,Benedetti:2015zsw,
  Biemans:2016rvp,Pagani:2016dof,Denz:2016qks,Falls:2017cze,
  Houthoff:2017oam,Knorr:2017fus}
to higher derivative and higher curvature extensions and variants thereof
\cite{Lauscher:2002sq,
  Codello:2006in,Codello:2007bd,Machado:2007ea,Codello:2008vh,
  Niedermaier:2009zz,Benedetti:2009rx,
  Rechenberger:2012pm,Benedetti:2012dx,Dietz:2012ic,
  Ohta:2013uca,Falls:2013bv,Benedetti:2013jk,Dietz:2013sba,
  Falls:2014tra,
  Eichhorn:2015bna,Ohta:2015efa,Ohta:2015fcu,
  Falls:2016wsa,Falls:2016msz,Gies:2016con,Christiansen:2016sjn,
  Gonzalez-Martin:2017gza,Becker:2017tcx}.
Strong quantum effects invariably modify the high-energy limit.
Interestingly, however, canonical mass dimension continues to be a good ordering principle \cite{Falls:2014tra}:
classically relevant couplings 
remain relevant while classically irrelevant couplings 
remain irrelevant \cite{Falls:2016msz}, including the notorious Goroff-Sagnotti term \cite{Gies:2016con}.  
Further  aspects such as diffeomorphism
invariance in the presence of a cutoff, and the r\^ole of background fields have also been clarified.

It then becomes natural to include matter fields, and to clarify the
impact of matter on asymptotic safety for gravity
\cite{Dou:1997fg,Percacci:2002ie,Narain:2009fy,Daum:2010bc,
  Folkerts:2011jz,FolkertsDiploma,Harst:2011zx,Eichhorn:2011pc,
  Eichhorn:2012va,Dona:2012am,Henz:2013oxa,Dona:2013qba,
  Percacci:2015wwa,Labus:2015ska,Oda:2015sma,Meibohm:2015twa,Dona:2015tnf,
  Meibohm:2016mkp,Eichhorn:2016esv,Henz:2016aoh,Eichhorn:2016vvy,
  Christiansen:2017gtg,Eichhorn:2017eht,Biemans:2017zca,Christiansen:2017qca,
  Eichhorn:2017ylw,Eichhorn:2017lry,Eichhorn:2017egq,Alkofer:2018fxj}.  In general it
is found that matter fields constrain asymptotic safety for gravity,
although not all specifics for this are fully settled yet.  In
expansions about flat backgrounds, it was noticed that the graviton
dominates over free matter field fluctuations, either via an
enhancement of the graviton propagator or the growth of the graviton
coupling \cite{Meibohm:2015twa}.  This pattern should play a r\^ole for
asymptotic safety of the fully coupled theory, and for weak gravity
bounds \cite{Eichhorn:2016esv,Christiansen:2017gtg,Eichhorn:2017eht}.
In a similar vein, the impact of quantised gravity on gauge theories
has been investigated within perturbation theory
\cite{Robinson:2005fj,Pietrykowski:2006xy,Toms:2007sk,Ebert:2007gf,Tang:2008ah,Toms:2010vy}
by treating gravity as an effective field theory
\cite{Donoghue:1993eb}, and within the asymptotic safety scenario
\cite{Daum:2010bc,Folkerts:2011jz,FolkertsDiploma}.  Modulo gauge and
scheme dependences, all studies find the same negative sign for the
Yang-Mills beta function $(\beta<0)$ in support of asymptotic freedom.
The reason for this was uncovered in
\cite{Folkerts:2011jz,FolkertsDiploma}: Due to an important
kinematical identity (\autoref{fig:kinID}), related to diffeomorphism 
and gauge invariance, $\beta <0$ follows automatically, and
irrespective of the gauge or regularisation.

In this paper, we want to understand the prospect for asymptotic
safety of quantum gravity coupled to matter. To that end, we combine
general, formal considerations with detailed and explicit studies
using functional renormalisation. A main new addition is a formal line
of reasoning, which explains why and how gravitons dominate the
high-energy behaviour, largely independently of the matter fields as
long as these remain sufficiently weakly coupled. Using functional
renormalisation, this is then put to work for $SU(N_c)$
Yang-Mills theory coupled to gravity.  In an expansion about flat
backgrounds, explicit results for beta functions, fixed points,
universal exponents, and scaling solutions are given. Systematic
approximations exploiting running propagators, the three-graviton and
the graviton-gauge vertices are performed up to including independent
couplings for gauge-gravity and pure gravity interactions, and for the
background couplings. Care is taken to distinguish fluctuating and
background fields.  Invariably, we find that the gauge coupling
becomes asymptotically free while the gravitational sector becomes
asymptotically safe.  The dependence on matter field multiplicities is
weak. We also investigate the scheme dependence, which is found to be
more pronounced, and explain how it can be handled without changing
the physics.  This allows us to offer a new interpretation of many
earlier results and to lift some of the tensions amongst previous
findings.

This paper is organised as follows. In \autoref{sec:as-af} we present
a formal argument for asymptotic safety of Yang-Mills--gravity, and
extensions to general matter--gravity systems. In \autoref{sec:frg} we
introduce the RG for Yang-Mills--gravity, and some
notation and conventions.  In \autoref{sec:QG_to_YM}, we analyse
whether asymptotic freedom in Yang-Mills theories is maintained when
coupled to a dynamical graviton. Conversely, in
\autoref{sec:YM_to_QG}, the influence of gluon fluctuations on
UV-complete theories for gravity are studied.  In \autoref{sec:ASYMG},
asymptotic safety of the fully-coupled Yang-Mills--gravity system is
investigated in the standard uniform approximation with a unique
Newton's coupling. We further discuss the stable large-$N_c$ limit of
this system.  In \autoref{sec:general}, we lift the uniform
approximation and discuss the system with separate Newton's couplings
for gauge-gravity and pure gravity interactions. We also discuss the
RG scheme dependence and relate our findings with earlier ones in the
literature.  In \autoref{sec:summary}, we briefly summarise our
findings. The Appendices comprise the technical details.

\section{From asymptotic freedom to asymptotic safety}
\label{sec:as-af}
In this section, we provide our main line of reasoning for why matter
fields, which are free or sufficiently weakly coupled in the UV --
such as in asymptotic freedom -- entail asymptotic safety in the full
theory including gravity. Throughout, Yang-Mills theory serves as the
principle example.

\subsection{Yang-Mills coupled to gravity: the setup}
Any correlation function approach to gravity works within an expansion
of the theory about some generic metric. The necessity of gauge fixing
in such an approach introduces a background metric into the approach.
Hence, we use a background field approach in the gauge sector, giving
us a setting with a combined background
$\bar g_{\mu\nu}, \bar A_\mu^a$. Background independence is then
ensured with the help of Nielsen or split Ward-Takahashi identities
and the accompanying Slavnov-Taylor identities (STIs) for both the
metric fluctuations and the gauge field fluctuations.  The superfield
$\phi$ comprises all fluctuations or quantum fields with
\begin{align}\label{eq:phi}
 A_\mu&=\bar A_\mu+a_\mu\,,\quad\quad g_{\mu\nu}=\bar g_{\mu\nu}+\sqrt{G}\,h_{\mu\nu}\,, \notag\\ 
\phi&=\left(h_{\mu\nu},c_\mu,\bar{c}_\mu,a_\mu,c,\bar{c}\right) \,, 
\end{align}
with the dynamical fluctuation graviton $h_{\mu\nu}$ and gauge field $a_\mu$. In
\eqref{eq:phi}, $c_\mu$ and $ {c}$ are the gravity and
Yang-Mills ghosts respectively.  The classical Euclidean action of the
Yang-Mills--gravity system is given by the sum of the gauge-fixed
Yang-Mills and Einstein-Hilbert actions, 
\begin{align}\label{eq:Scl} 
  S_{\text{\tiny{cl}}}[\bar g,\bar A; \phi] = S_{\text{\tiny{gauge}}}[\bar g,\bar A; \phi]  
  + S_{\text{\tiny{gravity}}} [\bar g,\bar A; \phi] \,, 
\end{align} 
where the two terms
$S_{\text{\tiny gauge}}= S_A+ S_{A,\text{\tiny gf}} +S_{A,\text{\tiny gh}}$
and
$S_{\text{\tiny gravity}} = S_{\text{\tiny EH}}+S_{g,\text{\tiny gf}} +S_{g,\text{\tiny gh}}$
are the fully gauge fixed actions of
Yang-Mills theory and gravity respectively. The Yang-Mills action
reads
\begin{align}\label{classical_YM_action}
  S_{A}[g,A] = & \012 \int \mathrm{d}^4 x
  \sqrt{\mathrm{det}\,g} \, g^{\mu \mu'} g^{\nu \nu'} \, {\rm tr}\, 
  F_{\mu' \nu'} F_{\mu \nu}  \,,
\end{align}
where the trace in \eqref{classical_YM_action} is taken in the
fundamental representation, and
\begin{align}\label{eq:Fmunu}
 F_{\mu\nu} =\0{i}{g_s } [D_\mu, D_\nu]\,,
\quad D_\mu=\partial_\mu -i g_s A_\mu\,,\quad {\rm tr}\, t^a t^b = \012\,.
\end{align}
The classical Yang-Mills action \eqref{classical_YM_action} only depends
on the full fields $g,A$ and induces gauge-field--graviton
interactions via the determinant of the metric as well as the Lorentz
contractions and derivatives. The gauge fixing is done in the
background Lorentz gauge $\bar D_\mu a_\mu =0$ with
$\bar D= D_\mu(\bar A) $. The gauge fixing and ghost terms read
\begin{align}\nonumber 
  S_{A,\text{\tiny gf}} ={}&\0{1}{2\xi} \int \mathrm{d}^4 x
                         \sqrt{\mathrm{det}\,\bar g} \,( 
                         \bar g^{\mu \nu} \bar D_\mu a_\nu)^2\,,  \\[1ex] 
  S_{A,\text{\tiny gh}}={}&  \int \mathrm{d}^4 x\sqrt{\mathrm{det}\, \bar g}\,
                        \, \bar g^{\mu \nu}\, \bar c 
                        \bar D_\mu D_\nu\, c\,, 
\label{eq:YMgaugefixing}
\end{align}
where we take the limit $\xi\to0$. The gauge fixing and ghost terms only depend on
the background metric and hence do not couple to the dynamical
graviton $h_{\mu\nu}$. The Einstein-Hilbert action is given by
\begin{align}\label{eq:EH_action}
  S_{\text{\tiny{EH}}} &= \0{1}{16 \pi G} \int \mathrm{d}^4 x \sqrt{\det g}
  \Bigl(2\Lambda-R(g)\Bigr)\,,
\end{align}
with a linear gauge fixing $F_\mu$ and the corresponding ghost term, 
\begin{align}\nonumber 
 S_{g,\text{\tiny gf}} =\01{2\alpha}\int\mathrm d^4x\,
  \sqrt{\det\bar g}\,\bar g^{\mu\nu}F_\mu F_\nu\,,\\[1ex] 
 S_{g,\text{\tiny gh}}=\int\mathrm d^4 x \,\sqrt{\det\bar{g}}\,  
\bar g^{\mu \mu'} \bar g^{\nu \nu'}\bar{c}_{\mu'}
  {\cal M}_{\mu\nu} c_{\nu'} \,.
\end{align} 
with the Faddeev-Popov operator ${\cal M}_{\mu\nu}(\bar g, h)$ of the gauge
fixing $F_\mu(\bar g, h)$. We employ a linear, de-Donder type gauge fixing,
\begin{align} \nonumber F_\mu ={}& \bar \nabla^\nu
  h_{\mu\nu} - \0{1+\beta}4 \bar \nabla_\mu {h^\nu}_\nu \,, \\[1ex] 
{\cal M}_{\mu\nu} ={}&
  \bar\nabla^\rho\left(g_{\mu\nu}\nabla_\rho+g_{\rho\nu}\nabla_\mu
  \right)-\bar\nabla_\mu\nabla_\nu\,, 
\label{eq:gravitygaugefixing} \end{align}
with $\beta=1$ and the limit $\alpha\to0$, which is a fixed point
of the ()RG flow \cite{Litim:1998qi}. 

\subsection{Asymptotic freedom in Yang-Mills with gravity}
\label{sec:ASgluegrav}
Gauge theories with gauge group $U(N)$ or $SU(N)$ describe the
electroweak and the strong interactions, and form the basis of the
Standard Model of particle physics.  A striking feature of non-Abelian
gauge theories is asymptotic freedom, meaning that the theory is
governed by a Gaussian fixed point in the ultraviolet, which implies
that gluon interactions weaken for high energies and that perturbation
theory is applicable.  In fact, the great success of the Standard
Model is possible only due to the presence of such a Gaussian fixed
point, which allows us to neglect higher order operators in the high
energy limit.  The weakening of interactions is encoded in the energy
dependence of the Yang-Mills coupling, which in turn is signalled by a
strictly negative sign of the beta function.  However, it is well
known that fermions contribute with a positive sign to the running of
the Yang-Mills coupling,
\begin{align}
\label{eq:QCD}
 \0{\beta_{\alpha_s}^{\text{\tiny{1-loop}}}}{\alpha_s^2} \equiv \mu\0{ \partial\alpha_s}{\partial\mu}\0{1}{\alpha_s^2}  
 = -\0{1}{4 \pi} \left(\0{22}{3}N_c - \0{4}{3} N_f\right) \,,
\end{align}
where we have displayed only the one-loop contributions with $N_c$ and
$N_f$ denoting the number of colours and fermion flavours, and
$\alpha_s={g_s^2}/(4 \pi)$. One can see that there is a critical
number of fermion flavours $N_f^{\rm crit}=\0{11}{2}N_c$ above which
the one-loop beta function changes sign.  This implies that asymptotic
freedom is lost.  It has been noted recently that gauge theories with
matter and without gravity may very well become asymptotically safe in
their own right
\cite{Litim:2014uca,Litim:2015iea,Bond:2016dvk,Bond:2017wut,Bond:2017lnq,Bond:2017suy}.

Returning to gravity, it has been shown in
\cite{Robinson:2005fj,Pietrykowski:2006xy,Toms:2007sk,Ebert:2007gf,Tang:2008ah,
  Toms:2010vy,Daum:2010bc,Folkerts:2011jz,FolkertsDiploma} that graviton fluctuations lead to an
additional negative term $\beta_{\alpha_s,h}$ in
$\beta_{\alpha_s}\to \beta_{\alpha_s,a}+\beta_{\alpha_s,h}$ where
$\beta_{\alpha_s,a}$ is the pure gauge theory contribution
\eqref{eq:QCD}. The graviton contribution has a negative sign,
\begin{align}\label{eq:afassist}
\beta_{\alpha_s,h} \leq 0\,.  
\end{align}
Because of the lack of perturbative renormalisability this term is gauge-
and regularisation-dependent.  However, it has been shown that it is
always negative semi-definite, \cite{Folkerts:2011jz,FolkertsDiploma},
based on a kinematic identity related to diffeomorphism
invariance. Hence, asymptotic freedom in Yang-Mills theories is
assisted by graviton fluctuations. In the case of $U(1)$, they even
trigger it. This result allows us to already get some insight into the
coupled Yang-Mills--gravity system within a semi-analytic
consideration in an effective theory spirit: In the present work we
consider coupled Yang-Mills--gravity systems within an expansion of
the pure gravity part in powers of the curvature scalar as well as
taking into account the momentum dependence of correlation
functions. In the Yang-Mills sub-sector we consider an expansion in
$\tr\,F^n$ and $(\tr F^2)^n$, the lowest non-classical terms being
\begin{align}\label{eq:w2v4def}
w_2\left(\tr F^2\right)^2\,,\qqquad  v_4 \tr F^4\,.
\end{align}
Asymptotic freedom allows us to first integrate out the gauge
field. This sub-system is well-described by integrating out the gauge
field in a saddle point expansion within a one-loop
approximation. Higher loop orders are suppressed by higher powers in
the asymptotically free gauge coupling. This leads us to the effective
action
\begin{align}\nonumber 
  \Gamma[\bar g,\bar A, \phi]={}&S_{\text{\tiny{gravity}}}[\bar g; \phi] 
  + S_{\text{\tiny{gauge}}}[\bar g,\bar A; \phi]\\[1ex]
  &\hspace{-.6cm}-\012 \Tr \ln \left[\Delta_1\delta_{\mu\nu} 
    +\left(1-\01\xi\right) \nabla_\mu \nabla_\nu 
    \right]_{k_{a}^{\text{\tiny IR}}}^{k_{a}^{\text{\tiny UV}}}\,, 
 \label{eq:intA}
\end{align}
where $\Delta_1$ represents the spin-one Laplacian and
$k_{a}^{\text{\tiny IR}},k_{a}^{\text{\tiny UV}}$ indicate
diffeomorphism-preserving infrared and ultraviolet regularisations of
the one-loop determinant. Most conveniently this is achieved by a
proper-time regularisation, for a comprehensive analysis within the
FRG framework see \cite{Litim:2002xm,Litim:2002hj}. In any case, both
regularisations depend on the metric $g_{\mu\nu}$ and the respective scales
$k_{a}^{\text{\tiny IR}},k_{a}^{\text{\tiny UV}}$. The computation
can be performed with standard heat-kernel methods. 

The infrared sector of the theory is not relevant for the present
discussion of the fate of asymptotic safety in the ultraviolet. Note
also that Yang-Mills theory exhibits an infrared mass gap with the
scale $\Lambda_{\text{\tiny QCD}}$ due to its confining dynamics. In
covariant gauges as used in the present work this mass gap results in
a mass gap in the gluon propagator, for a treatment within the current
FRG approach see \cite{Cyrol:2016tym,Cyrol:2017ewj} and references
therein. This dynamical gaping may be simulated here by simply
identifying the infrared cutoff scale with
$\Lambda_{\text{\tiny QCD}}$.

Moreover, even though integrating out the gauge field generates higher
order terms such as \eqref{eq:w2v4def} in the UV, they are suppressed by
both, powers of the UV cutoff scale as well as the asymptotically free
coupling. Accordingly, we drop the higher terms in the expansion of
the Yang-Mills part of the effective action \eqref{eq:intA}. Note that
they are present in the full system as they are also generated by
integrating out the graviton. This is discussed below.

It is left to discuss the pure gravity terms that are generated by
ultraviolet gluon fluctuations in \eqref{eq:intA}. They can be expanded
in powers and inverse powers of the UV-cutoff scale
$k_a=k_{a}^{\text{\tiny UV}}$. This gives an expansion in powers of
the Ricci scalar $R$ and higher order invariants.
From the second line of \eqref{eq:intA} we are led to
\begin{align}
  & (N_c^2-1) \Biggl[  c_{g,a} k_{a}^2\int \mathrm{d}^4 x \sqrt{\det\, g}\,
    \Bigl(2 c_{\lambda,a} k_{a}^2- R\Bigr) \notag \\[1ex] 
  & +c_{R^2,a}  \int \mathrm{d}^4 x 
    \sqrt{\det\, g}\,\left(   R^2 + 
    z_{a} R_{\mu\nu}^2\right)\ln\0{R+{k_a^{\text{\tiny IR}}}^2}{k_a^2}\Biggr]  \notag \\[1ex] 
&  + \CO\left(\0{R^3}{k_a^2}\right)\,, 
 \label{eq:Aexpand}
\end{align}
where we suppressed potential dependences on $\Delta_g$ and
$\nabla_\mu$, in particular in the logarithmic terms. The
logarithm also could contain further curvature invariants such as
$R_{\mu\nu}^2$. In the spirit of the discussion of the confining
infrared physics we may substitute
$k_a^{\text{\tiny IR}}\to \Lambda_{\text{\tiny{QCD}}}$ in a full
non-perturbative analysis. In \eqref{eq:Aexpand}, the coefficients
$c_{g,a}, c_{\lambda,a}, c_{R^2,a}$ and $z_a$, are
regularisation-dependent and lead to contributions to Newton's
coupling, the cosmological constant, as well as generating an
$R^2$-term and potentially an $R_{\mu\nu}^2$ term. In the present
Yang-Mills case, $c_{g,a}$ is positive for all regulators. For fermions
and scalars, the respective coefficients $c_{g,\psi}, c_{g,\phi}$ are
negative. In summary, this leaves us with an asymptotically free
Yang-Mills action coupled to gravity with redefined couplings
\begin{align}
  G_{\text{\tiny eff}}&=\0{G}{1+(N_c^2-1) c_{g,a}  k_{a}^2 G}\,,\notag \\
 \0{ \Lambda_{\text{\tiny eff}} }{G_{\text{\tiny eff}}}&= \0{\Lambda}{G} + 
 (N_c^2-1) c_{g,a} 
c_{\lambda,a}k_{a}^4  \,. 
\label{eq:Nc-redef}\end{align} 
The coupling parameters $G,\Lambda$ should be seen as bare couplings
of the Yang-Mills--gravity system and chosen such that the
(renormalised) couplings
$G_{\text{\tiny eff}},\Lambda_{\text{\tiny eff}}$ are
$k_{a}$ independent. This corresponds to a standard renormalisation
procedure (introducing the standard RG scale $\mu_{\text{\tiny{RG}}}$) and
leads to $G(N_c,k_a),\Lambda(N_c,k_a)$. Note that demanding
$k_a$ independence of the effective couplings also eliminates their
$N_c$ running. For example, for the effective Newton's coupling
\begin{align}\label{eq:Nc-ka--scaling}
  ( N^2_c-1) \partial_{(N^2_c-1)}  \ln G_{\text{\tiny eff}}= 
  k_a^2 \partial_{k_a^2}  
  \ln G_{\text{\tiny eff}} = 0 \,, 
\end{align} 
holds in a minimal subtraction scheme where the
renormalisation scale $\mu_{\text{\tiny{RG}}}$ does not introduce
further $N_c$-dependencies, most simply done with
$\mu_{\text{\tiny{RG}}}$-independent couplings $G,\Lambda$.

We also have to include $ g_{R^2} R^2$  and
$ g_{R_{\mu\nu}^2} R_{\mu\nu}^2$ terms in the classical gravity action
in order to renormalise also these couplings,
\begin{align}
  g_{R^2,\text{\tiny eff}}&=g_{R^2} +(N_c^2-1) c_{R^2,a} 
      \ln\0{{k_a^{\text{\tiny IR}}}^2}{k_a^2} \,,\notag \\
  g_{R_{\mu\nu}^2,\text{\tiny eff}}&=g_{R_{\mu\nu}^2} +(N_c^2-1) 
      c_{R^2,a} z_a  \ln\0{{k_a^{\text{\tiny IR}}}^2}{k_a^2} \,.
\label{eq:Nc-redef-R^2}
\end{align}
Here, the minimal subtraction discussed above requires
$g_{R^2}(N_c, \ln k_a/k_a^{\text{\tiny IR}})$ and
$g_{R_{\mu\nu}^2}(N_c, \ln k_a/k_a^{\text{\tiny IR}})$. This leaves us with
a theory, which includes all ultraviolet quantum effects of the
Yang-Mills theory. Accordingly, in the ultraviolet its effective
action \eqref{eq:intA} resembles the Einstein-Hilbert action coupled
to the classical Yang-Mills action with appropriately redefined
couplings. It also has $R^2$  and $R_{\mu\nu}^2$ terms. However, the
latter terms are generated in any case by graviton fluctuations so
there is no structural difference to standard gravity with the
Einstein-Hilbert action coupled to the classical Yang-Mills.  

The only relevant $N_c$ dependence originates in the logarithmic
curvature dependence of the marginal operators $R^2$ and
$R_{\mu\nu}^2$ leading e.g.\ to
\begin{align}\label{eq:R2logR}
  \left( N_c^2-1 \right)  c_{R^2,a}\int \mathrm{d}^4 x 
  \sqrt{\det \, g}\,   R^2  \ln\left(  1+ \0{R}{ {k_a^{\text{\tiny IR}}}^2}\right) \,. 
\end{align}
These terms are typically generated by flows towards the infrared, for
a respective computation in Yang-Mills theory see
\cite{Eichhorn:2010zc}. Such a running cannot be absorbed in the pure
gravity part without introducing a non-local classical action.  From
its structure, the logarithmic running in \eqref{eq:Nc-redef-R^2}
resembles the one of the strong coupling in many flavour QCD: the
r\^ole of the gravity part here is taken by the gluon part
in many flavour QCD and that of the Yang-Mills part here is taken by
the many flavours. Accordingly, a fully conclusive analysis has to take
into account these induced interactions. This is left to future work,
here we concentrate on the Einstein-Hilbert part. The respective
truncation to matter-gravity systems have been studied at length in
the literature, and the arguments presented here fully
apply. Note also that the current setup (and the results in the
literature) can be understood as a matter-gravity theory, where the
respective terms are removed by an appropriate classical gravity
action that includes, e.g.,  $R^2 \ln R$ terms.
The discussion of these theories is also  linked to
the question of unitary in asymptotically safe gravity.

If we do not readjust the effective couplings within the minimal
subtraction discussed above they show already the fixed point scaling
to be expected in an asymptotically safe theory of quantum gravity,
see \eqref{eq:Nc-redef} and \eqref{eq:Nc-redef-R^2}. This merely reflects
the fact that Yang-Mills theory has no explicit scales. If we only
absorb the $k_{a}$ running of the couplings while leaving open a
general $\mu_{\text{\tiny{RG}}}$ dependence, the effective Newton's
coupling $G_{\text{\tiny eff}}$ scales with $1/N_c^2$, while the
effective cosmological constant scales with $N_c^0$.

In any case we have to use $G_{\text{\tiny eff}}$ for the gravity
scale in the Yang-Mills--gravity system instead of $G$.  For example, the
expansion of the full metric $g_{\mu\nu}$ in a background and a fluctuation then
reads
\begin{align}\label{eq:gheff}
g_{\mu\nu} = \bar g_{\mu\nu} + \sqrt{ G_{\text{\tiny eff}}} \,h_{\mu\nu}\,, 
\end{align}
with the dimension-one field $h_{\mu\nu}$ in the $d=4$ dimensional
Yang-Mills--gravity system. 

\subsection{Asymptotic safety in gravity with Yang-Mills}
It is left to integrate out graviton fluctuations on the basis of the
combined effective action, where the pure gravity part is of the
Einstein-Hilbert type. The couplings of the pure gravity sector, in
particular, Newton's coupling and the cosmological constant only receive
quantum contributions from pure gravity diagrams, while pure gauge and
gauge-graviton couplings only receive contributions from diagrams that
contain at least one graviton line. This system is
asymptotically safe in the pure gravity sector and assists asymptotic
freedom for the minimal gauge coupling, see \eqref{eq:QCD} and
\eqref{eq:afassist}, and leads to graviton-induced higher-order coupling
such as \eqref{eq:w2v4def}. In summary, we conclude that Yang-Mills--gravity
systems are asymptotically safe. The flow of this system and its
completeness is discussed in \autoref{sec:general}. 

The present analysis is also
important for the evaluation of general matter-gravity systems: we
have argued that asymptotic freedom of the Yang-Mills theory allows us
to successively integrate out the degrees of freedom, starting first
with the Yang-Mills sector. Evidently, this is also true for matter-gravity
systems with free matter such as treated comprehensively, e.g., in
\cite{Dona:2013qba} and \cite{Meibohm:2015twa}. In the former,
fermions and scalars were found to be unstable for a large flavour numbers
while in the latter fermions were shown to be
stable. For scalars, the situation was inconclusive as the anomalous
dimension of the graviton was exceeding an upper bound, $\eta_h<2$,
beyond which a regulator of the form
$R_{h,k}(p^2) \propto Z_{h} R^{(0)}_{h,k}(p^2)$ with
$R^{(0)}_{h,k}(0) =k^2$ is no longer a regulator with the cutoff scale $k$, 
\begin{align}\label{eq:nonReg}
  \lim_{k\to \infty}  R_{h,k}(0)\propto (k^2)^{1-\eta_h/2}\to 0\,,\quad {\rm for}\quad \eta_h>2\,. 
\end{align}
This bound can be pushed to $\eta_h<4$ but also this bound was
exceeded, see \cite{Meibohm:2015twa}. While the differences in the
stability analysis can be partially attributed to the different
approximations in \cite{Dona:2013qba} and \cite{Meibohm:2015twa}
(the former does not resolve the difference between background
gravitons and fluctuation gravitons in the pure gravity sector), we
come to conclude here, that both (and all similar ones) analyses lack
the structure discussed above. This calls for a careful reassessment
of the UV flows of matter-gravity systems also in the view of
relative cutoff scales. The latter is since long a well-known problem
in quantum field theoretical applications of the FRG, in particular,
in boson-fermion systems. For example, in condensed matter systems it
has been observed that exact results for the three-body scattering
(STM), see \cite{Diehl:2007xz}, can only be obtained within a
consecutive integrating out of degrees of freedom in local
approximations. If identical cutoff scales are chosen, the three-body
scattering only is described approximately. For a recent analysis of
relative cutoff scales in multiple boson and boson-fermion systems,
see \cite{Pawlowski:2015mlf}. 

In summary, the gravitationally coupled free-matter--gravity systems,
Yang-Mills--gravity systems, or more generally asymptotically free
gauge-matter--gravity systems are asymptotically safe, independent of
the number of matter degrees of freedom if this holds for one degree
of freedom or more generally if this holds for the minimal number of
degrees of freedom that already has the most general interaction
structure of the coupled theory. Phrased differently: simple large
$N$ scaling cannot destroy asymptotic safety, with $N$ being the
number of gauge-matter degrees of freedom.  

We emphasise that the analysis of such a minimal system as defined
above is necessary. It is not sufficient to rely on the fact that the
matter or gauge part can be integrated out first as gravity
necessarily induces non-trivial matter and gauge self-interactions at
an asymptotically safe gravity fixed point 
\cite{Eichhorn:2011pc,Eichhorn:2012va,Meibohm:2016mkp,Christiansen:2017gtg,Eichhorn:2017eht}.
If these self-interactions do not destroy asymptotic safety,
the systems achieve asymptotic safety for a general number of matter
or gauge fields by guaranteeing the ultraviolet dominance of graviton
fluctuations.

With these results at hand, we can now ask the question whether a 
"relative scaling" of  gravity vs matter cutoffs maintains the observed graviton dominance.
A natural "scaling hierarchy" for the cutoff scales
$k_h$ in the gravity and  $k_a$ in the Yang-Mills sector is motivated by 
the following heuristic consideration:
while gravity
feels the effective Newton's coupling $G_{\text{\tiny eff}}$, and
hence, graviton fluctuations and gravity scales should be measured in
$G_{\text{\tiny eff}}$, the Yang-Mills field generates contributions
to the (bare) Newton's coupling $G$. Assuming that both are of a similar strength, this leads  to 
\begin{align}\label{eq:kh-ka}
G_{\text{\tiny eff}}\,k_h^2\simeq  G\, k^2_a 
\end{align}
for the respective cutoff scales. Interestingly though, under this  hierarchy of scales, the
$N_c$ dependence of the coupled system disappears, and, within an
appropriate fine-tuning of the relation \eqref{eq:kh-ka}, the fixed point
values of Newton's coupling and the cosmological constant show no
$N_c$ dependence at all.  Stated differently, a rescaling such as in \eqref{eq:kh-ka} 
 guarantees
the dominance of graviton fluctuations over gauge or matter
fluctuations as long as the gauge-matter system is
asymptotically free. The phenomenon of graviton dominance as observed with identical cutoffs continues 
to be observed under  a weighted rescaling \eqref{eq:kh-ka}.

We close this chapter with some remarks. 
\begin{itemize} 
\item[(1)]The naturalness of the rescaling \eqref{eq:kh-ka} is finally
  decided by taking into account momentum or spectral dependencies of
  the correlation functions. This is at the root of the question of
  stability and instability of matter-gravity systems. It is
  here where the marginal, logarithmically running, terms such as 
  \eqref{eq:R2logR} come into play. They are not affected by this
  rescaling, which also shows their direct physics relevance.
\item[(2)] Within the above rescaling, the fixed point of the
  gravity-induced gauge couplings such as $w_2$ and $v_4$, see
  \eqref{eq:w2v4def}, are of order ${g^*}^4$ of the pure gravity fixed point
  coupling $g^*$. Note however, that this value can be changed by
  readjusting the rescaling \eqref{eq:kh-ka}. 
\item[(3)] Note that within the dynamical re-adjustment of the scales
  the fixed point Newton's coupling gets weak, $g^*\propto 1/N_c^2$. In
  other words, gravity dominates by getting weak. This is in line with
  the weak-gravity scenario advocated recently 
  \cite{Eichhorn:2016esv,Christiansen:2017gtg,Eichhorn:2017eht}.
  However, its physical foundation is different. 
\item[(4)] For a sufficiently large truncation, the theory should
  be insensitive to a relative rescaling of the cutoff scales
  $k_{\text{\tiny{gravity}}}$ and $k_{\text{\tiny{matter}}}$ and to
  other changes of the regularisation scheme. This is partially investigated
  in \autoref{sec:general}. Moreover, in all of the following
  RG computations we do not resort to the rescaling \eqref{eq:kh-ka} 
  but use identical cutoff scales $k_{\text{\tiny{gravity}}}=k_{\text{\tiny{matter}}}$.
\end{itemize}  
In the following analysis, we will refer to the present chapter for an
evaluation of our results.

\section{Renormalisation group}
\label{sec:frg}
In the present work, we quantise the Yang-Mills--gravity system within
the functional renormalisation group (FRG) approach. The general idea is to
integrate-out quantum fluctuations of a given theory successively,
typically in terms of momentum or energy shells, $p^2 \sim k^2$. This
procedure introduces a scale dependence of the correlation functions,
which is most conveniently formulated in terms of the scale-dependent
effective action $\Gamma_k$, the free energy of the theory. Its
scale-dependence is governed by the flow equation for the effective
action, the Wetterich equation \cite{Wetterich:1992yh},
see also \cite{Ellwanger:1993mw,Morris:1993qb},
\begin{align}
\partial_t \Gamma_k[\bar g;\phi] = \012 \, \Tr
 \left[\0{1}{ \Gamma^{(0,2)}_k[\bar g;\phi] +
  R_k} \,\partial_t R_k\right]\,,  
\label{eq:WettEq}
\end{align} 
where the trace sums over species of fields, space-time, Lorentz,
spinor, and gauge group indices, and includes a minus sign for
Grassmann valued fields.  For the explicit computation, we employ the
flat regulator \cite{Litim:2000ci,Litim:2001up}, see
App.~\ref{app:regulators}.  From here on, we drop the index $k$ for
notational convenience.  The scale dependence of couplings, wave
function renormalisations, or the effective action is implicitly
understood.

The computation utilises the systematic vertex expansion scheme as presented in
\cite{Christiansen:2012rx,Christiansen:2014raa,Christiansen:2015rva,
  Meibohm:2015twa,Christiansen:2016sjn,Denz:2016qks} for pure gravity as well as
matter-gravity systems: the scale dependent effective action that
contains the graviton-gluon interactions is expanded in powers
of the fluctuation super field $\phi$ defined in \eqref{eq:phi},
\begin{align}
\Gamma[\bar{g},\bar A;\phi] = \sum_n \0{1}{n!}
\Gamma^{(\phi_{1}...\phi_{n})}_{\bf a_1 ... \bf a_n}[\bar{g},\bar A, 0]
\phi_{\bf a_1}...\phi_{\bf a_n} \,.
\label{eq:Gvertexp}
\end{align}
In \eqref{eq:Gvertexp}, we resort to de-Witt's condensed notation.
The bold indices sum over species of
fields, space-time, Lorentz, spinor, and gauge group indices. The
auxiliary background field is general. Here, we choose it as
$ \bar{\phi}=\left(\bar{A} = 0,\bar{g} = \mathbb{1} \right)$ for
computational simplicity.  In this work, we truncate such that we 
obtain a closed system of flow equations for
the gluon two- and the graviton two- and three-point
functions, $\partial_t \Gamma^{(aa)}$,
$\partial_t \Gamma^{(hh)}$, and $\partial_t \Gamma^{(hhh)}$. The
corresponding flow equations are derived from \eqref{eq:WettEq} by
functional differentiation.

The pure gravity part of the effective action
$\Gamma_{\mathrm{grav}}$ in \eqref{eq:Gvertexp} is constructed exactly
as presented in
\cite{Christiansen:2012rx,Christiansen:2014raa,Christiansen:2015rva,
  Meibohm:2015twa,Christiansen:2016sjn,Denz:2016qks}. This
construction is extended to the Yang-Mills part. Moreover, for the
flow equations under consideration here, only terms with at most two
gluons contribute. In summary, our approximation is based solely on the
classical tensor structures
$S_{\text{\tiny{cl}}}$ that are derived
from \eqref{eq:Scl}. The correlation functions follow as,
\begin{align}\label{eq:Gamman}
  \Gamma^{(\phi_{1}\ldots\phi_{n})}_{{\bf a_1}\ldots \bf{a_n}} =
  &\left(\prod_{i=1}^n Z^{\012}_{\phi_{i}} \right)
    S_{\text{\tiny{cl}},{\bf a_1}\ldots \bf{a_n}}^{(\phi_{1}\ldots\phi_{n})}
    \left({\bf p};g_{\phi_{1}\ldots\phi_{n}},\lambda_{\phi_{1}\ldots\phi_{n}}\right)\,,
\end{align}
where the $Z_{\phi_{i}}$ are the wave function renormalisations of the
corresponding fields and ${\bf p}=(p_1,...,p_n)$. The
$g_{\phi_{1}\cdots\phi_{n}},\lambda_{\phi_{1}\ldots\phi_{n}}$ are the
couplings in the classical tensor structures that may differ for each
vertex. In the present approximation, these couplings are
extracted from the momentum dependence at the symmetric point, and
hence, carry part of the non-trivial momentum dependence of the
vertices. The projection procedure is detailed later. We further
exemplify the couplings at the example of the pure graviton and the
gauge-graviton vertices. Each graviton $n$-point function,
$\Gamma^{(h_1\cdots h_n)}$, depends on the dimensionless parameters
\begin{subequations}\label{eq:dimless}
\begin{align}
  g_n \equiv g_{h^n}= G_n k^2\,,\quad 
  \lambda_n \equiv\lambda_{h^n}= \Lambda_n/k^2\,, 
  \label{eq:dimlessgh}
\end{align} 
and a mixed gauge-graviton $(n+2)$-point function on 
\begin{align}
  g_{a^2  h^n}=  G_{a^2  h^n} k^2\,, \qquad 
  g_{A^2  h^n}=  G_{A^2  h^n} k^2\,. 
  \label{eq:dimlessgA}
\end{align} 
\end{subequations}
In particular the parameters
$\lambda_n$ should not be confused with the cosmological constant, for
more details see, e.g., \cite{Denz:2016qks}. In the present approximation we
identify all gravity couplings 
\begin{align}\label{eq:gn}
  g_{A^m h^n}= g_3=:g\,,\quad \lambda_{n>2} = \lambda_3\,, 
  \quad \lambda_2 = -\012 \mu\,, 
\end{align}
the general case without this identification is discussed in \autoref{sec:general}. 
Note that the identification in \eqref{eq:gn} introduces (maximal)
diffeomorphism invariance to the effective action: in order to
eludicate this statement, we discuss the full effective action for
constant vertices. With $g=\bar g +\sqrt{G} Z^{1/2}_h h$ and
$A= \bar A +Z^{1/2}_a a$ and \eqref{eq:gn}, the current approximation 
can schematically be written as a sum of the classical action and a
mass-type term for the fluctuation graviton, 
\begin{align} \label{eq:diff+kin} 
  \Gamma[\bar g,\bar A;\phi]={}& 
  \left. S_{\text{\tiny{cl}}}[g,A] \right|_{G=G_3,\Lambda=\Lambda_3}
  +\Delta\Gamma[\bar g] \notag\\[1ex]
  &\,+\0{k^4}2 Z_h (\mu+2 \lambda_3)\, 
  h_{\bf a} \CT_{{\bf a}\bf{b}} h_{\bf b} \,,
\end{align}
where
$\CT_{\bf ab} =S_{\text{\tiny{EH}}\ {\bf a}\bf{b}}^{(hh)}
(p^2=0;g=1,\lambda=1)$ is the tensor structure of the second
derivative of the cosmological constant term.  The $\lambda_3$ term
cancels with the corresponding contribution in the first line, and
thus, $\mu$ is the coupling of this tensor structure.  This
is the minimal approximation that is susceptible to the non-trivial
symmetry identities, both the modified STIs and the Nielsen 
identities present in gauge-fixed quantum
gravity. This information requires the non-trivial running of wave
function renormalisations $Z_{\bar g}, Z_{\bar A},Z_h,Z_c,Z_a$, that
of the graviton mass parameter $\mu$, as well as the dynamical
gravity interactions $g$ and $\lambda_3$.  Note that at a (UV)
fixed point the flows of the couplings $\mu$, $g$, and
$\lambda_3$ vanish while the anomalous
dimensions do not vanish. 

The last identification in \eqref{eq:gn} reflects the fact that
$-2 \lambda_2 $ is the dimensionless mass parameter of the
graviton. Note however that $\mu$ is not a physical mass of the
graviton in the sense of massive gravity: in the classical regime of
gravity, it is identical to the cosmological constant,
$\bar\lambda=-\012 \mu$. Higher order operators in particular
$g_{a^n}$ may couple back in an indirect fashion, see, e.g.,
\cite{Christiansen:2017gtg}. 
In summary, this leads us to an expansion
of the mixed fluctuation terms (with both, powers of $a$ and powers of
$h$) of the effective action \eqref{eq:Gvertexp}
\begin{align}
 \label{eq:trunc_YM_grav}
 \Gamma&[\bar{g},\bar A; \phi] \Big|_{\text{\tiny{mixed}}}
 = \Gamma^{(ah)}_{\bf a_1 \bf a_2}\,a_{\bf a_1}  h_{\bf a_2} 
 + \0{1}{2} \Gamma^{(ahh)}_{\bf a_1 \bf a_2 \bf a_3} \,a_{\bf a_1}  h_{\bf a_2} h_{\bf a_3}
 \notag\\ 
 &+   \012 \Gamma^{(aah)}_{ \bf a_1 \bf a_2 \bf a_3}\,a_{\bf a_1} a_{\bf a_2} h_{\bf a_3} 
 +  \0{1}{4} \Gamma^{(aahh)}_{ \bf a_1 \bf a_2 \bf a_3 \bf a_4} \,a_{\bf a_1} a_{\bf a_2} h_{\bf a_3} h_{\bf a_4}
 \notag\\
 &+ \0{1}{12} \Gamma^{(aahhh)}_{ \bf a_1 \bf a_2 \bf a_3 \bf a_4 \bf a_5} \,a_{\bf a_1} a_{\bf a_2} h_{\bf a_3} h_{\bf a_4} h_{\bf a_5} 
 +\mathcal{O}\left(a^3 h,a h^3 \right) \, . 
\end{align}
As we consider also correlation functions of the background gluon, we
need the expansion of the fluctuation vertices in
\eqref{eq:trunc_YM_grav} in the background field, i.e.,
\begin{align}\label{eq:agA}
 \Gamma^{(ah)}_{
\bf a_1 \bf h_2} [\bar A]= \Gamma^{(ah)}_{
\bf a_1 \bf h_2}[0] + \Gamma^{(\bar Aah)}_{
\bf b_1 \bf a_1 \bf h_2}[0] \bar A_{\bf b_1}+O(\bar A^2)  \,,
\end{align}
in an expansion about vanishing background gauge field. In the
following, we consider trivial metric and gluon backgrounds
$\bar g=\id $ and $\bar A=0$. In this background, the terms of the
order $\mathcal{O}(a^3 h,a h^3)$ do not enter the flow equations of
the gluon and graviton propagators nor that of the graviton
three-point function. This is the reason why they have not been
displayed explicitly in \eqref{eq:trunc_YM_grav}. Note that with this background 
choice, the terms linear in $a$ in the second line in \eqref{eq:trunc_YM_grav} vanish.

In this trivial background,
we can use standard Fourier representations for our correlation
functions. In momentum space, the above correlation functions are given
as follows: the gluon two-point function reads
\begin{align}\label{eq:Gaa}
\Gamma^{(aa)}_{\mu \nu}(p_1,p_2) = 
Z_{a}^{\0{1}{2}}(p_1^2)
Z_{a}^{\0{1}{2}}(p_2^2) \left. \0{ \delta^2
S_A}{\delta {a}^{\mu}(p_1) \delta {a}^{\nu}(p_2)} 
\right|_{\phi=0}\, . 
\end{align}
The graviton two-point function is parameterised according to the
prescription presented in
\cite{Christiansen:2012rx,Christiansen:2014raa,Christiansen:2015rva,
  Meibohm:2015twa,Christiansen:2016sjn,Denz:2016qks},
\begin{align}\label{eq:Ghh}
\Gamma^{(hh)}_{\mu \nu \alpha \beta}(p_1,p_2) = Z_{h}^{\0{1}{2}}(p_1^2)
Z_{h}^{\0{1}{2}}(p_2^2) \left. \0{ G_2 \, \delta^2
S_{\text{\tiny{EH}}}(G_2,\Lambda_2)}{\delta {h}^{\mu \nu}(p_1) \delta {h}^{\alpha \beta}(p_2)} 
\right|_{\phi=0}\, ,
\end{align}
where $- 2 \Lambda_2  = \mu\,k^2$ as introduced in \eqref{eq:gn}. Note
that the right-hand side of \eqref{eq:Ghh} does not depend on $G_2$. The
two-gluon--one-graviton vertex is given by
\begin{align}\label{eq:Gaah}
\Gamma^{(aah)}_{\mu \nu \alpha \beta}(p_1,p_2,p_3)  &=
Z_a^{\0{1}{2}}(p_1^2)
Z_a^{\0{1}{2}}(p_2^2) Z_h^{\0{1}{2}}(p_3^2) 
\\ \notag &  
\left. \times \0{G^{\0{1}{2}}_3 \delta^3
S_A}{\delta a^{\mu}(p_1) \delta a^{\nu}(p_2) \delta
h^{\alpha \beta}(p_3) } \right|_{\phi=0} \, ,
\end{align}
with scale- and momentum-dependent wave function renormalizations
$Z_a$ for the gluon and $Z_h$ for the graviton and a scale-dependent
gravitational coupling $G_3$.
The other $n$-point functions have a completely analogous construction,
which is not displayed here.

In addition to the fluctuation vertices, we also need mixed vertices
involving two background gluons and the fluctuation fields as in
\eqref{eq:agA}, $\Gamma^{A^2 h^n}$ and $\Gamma^{A^2 a^n}$ with
$n=1,2$. They are parameterised as in \eqref{eq:Gaa} - \eqref{eq:Gaah} with
$Z_a\to Z_A$. We also would like to emphasise two structures that
facilitate the present computations:
\begin{itemize}
\item[(1)] As we consider the flow equations for the gluon two-point function,
  and the graviton two- and three-point functions,
  only the terms quadratic in $a_\mu$ in \eqref{eq:trunc_YM_grav} contribute to the
  graviton-gluon interactions in the flow equations.
  The non-Abelian parts in the $F^2$ term do not
  contribute since they are of order three and higher. Hence,
  modulo trivial colour factors $\delta^{ab}$, the vertices
  defined above are identical for $SU(N)$ and $U(1)$ gauge theories.
\item[(2)] In principle, the derivatives in $F^{\mu \nu}$ are covariant
  derivatives with respect to the Levi-Civita connection. However,
  since $F^{\mu \nu}$ is asymmetric, and the Christoffel-symbols symmetric in
  the paired index, the latter cancel out, and the covariant
  derivatives can be replaced by partial derivatives. 
\end{itemize}
In the end, we are interested in the gravitational corrections to the
Yang-Mills beta function, and the Yang-Mills contributions to the
running in the gravity sector. The beta functions of the
latter have been discussed in great detail in
\cite{Christiansen:2012rx,Christiansen:2014raa,Christiansen:2015rva,Christiansen:2016sjn,Denz:2016qks}.
In the Yang-Mills sector, we make use of the fact that the wave function
renormalisation $Z_A$ of the background gluon is related to the background
(minimal) coupling by
\begin{align}\label{eq:ZAZg}
Z_{\alpha_s} = Z_{A}^{-1} \,,  
\end{align}
which is derived from background gauge invariance of the
theory. The latter can be related to quantum gauge invariance with
Nielsen identities, see
\cite{Freire:2000bq,Litim:2002ce,Pawlowski:2003sk,Pawlowski:2005xe,Donkin:2012ud}
in the present framework. This also relates the background minimal
coupling to the dynamical minimal coupling of the fluctuation
field. Note that this
relation is modified in the presence of the regulator, in particular,
for momenta $p^2 < k^2$. There the interpretation of the
background minimal coupling requires some care. 
The running of the background coupling is then determined by
\begin{align}\label{eq:beta_YM}
\partial_t \alpha_s= \beta_{\alpha_s}= \eta_A\alpha_{s}\,, 
\end{align}
with the gluon anomalous dimension
\begin{align}
\eta_A := - \0{\partial_t Z_A}{Z_A} \,. 
\end{align}
Note that in general all these relations carry a momentum dependence
as $Z_A(p^2)$ carries a momentum dependence. This will become
important in the next section for the physics interpretation of the
results.

\begin{figure}[t]
\includegraphics[width=\linewidth]{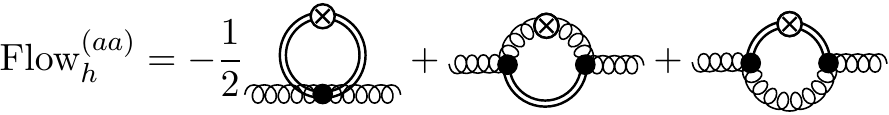}
\caption{Diagrammatic depiction of graviton contributions to the flow of
the gluon propagator. Wiggly and double lines represent gluon and 
graviton propagators, respectively.}
\label{fig:flow_grav_to_YM}
\end{figure}

\section{Graviton contributions to Yang-Mills}
\label{sec:QG_to_YM}
In this section we compute the gravitational corrections to
the running of the gauge coupling. The key question is if 
graviton-gluon interactions destroy or preserve the property of
asymptotic freedom in the Yang-Mills sector. 
The running of the gauge
coupling can be calculated from the background gluon wave function
renormalisation.  Its flow equation is derived from \eqref{eq:WettEq} with two functional 
derivatives w.r.t.\ $\bar A$. Schematically it reads  
\begin{align}
\partial_t \Gamma^{(\bar A\bar A)}(p) = \mathrm{Flow}^{(\bar A\bar A)}_A(p) +
\mathrm{Flow}^{(\bar A\bar A)}_h(p) \, , 
\end{align}
where the first term contains only gluon fluctuations and the second
term is induced by graviton-gluon interactions. The diagrammatic form
of the second term is displayed in \autoref{fig:flow_grav_to_YM}. This
split is reflected in a corresponding split of the anomalous dimension
\begin{align}
\eta_A(p^2) = \eta_{A,A}(p^2) + \eta_{A,h}(p^2) \,. 
\end{align}
Note that in the present approximation we have
$\eta_{A,h}=\eta_{a,h}$. This originates in the fact that
the fluctuation graviton only couples to gauge invariant operators.

Asymptotic freedom is signalled by a negative sign of the gluon
anomalous dimension as the beta function for the coupling is
proportional to $\eta_A$. We know that the pure gluon contributions
$\eta_{A,A}$ are negative. Hence, the question whether asymptotic
freedom is preserved in the Yang-Mills--gravity system boils down to
the sign of the gravity contributions $\eta_{A,h}$, and we arrive at
\begin{align}\label{eq:as}
\eta_{A,h} \leq 0 \quad \Longleftrightarrow
\quad {\rm asymptotic\ freedom} \, . 
\end{align}
The anomalous dimension in \eqref{eq:as} depends on cutoff and
momentum scales. For small momentum scales $p^2/k^2 \to 0$ the
regulator induces a breaking of quantum-gauge and
quantum-diffeomorphism invariance: the respective STIs of the
fluctuation field correlation functions are modified. This
necessitates also a careful investigation of the background
observables, which only carry physics due to the relation of background
gauge- and diffeomorphism invariance. 

Note that asymptotic freedom as defined in \eqref{eq:as} only applies to
the minimal coupling. Higher order fluctuation couplings are not
necessarily vanishing. Indeed, it has been shown that the
asymptotically safe fixed points of general matter and gauge fields
coupled to gravity can not be fully asymptotically free in the matter
and gauge field sector, see
\cite{Eichhorn:2012va,Meibohm:2016mkp,Eichhorn:2016esv,Christiansen:2017gtg,Eichhorn:2017eht}.
In the present work, this leads to $a^4$ vertices from higher-order
invariants such as $({\rm tr} F^2)^2$ and ${\rm tr} F^4$ with fixed
point values proportional to $g_a^2/(1 +\mu)^3$ with $g_a=g$ in our
approximation. Moreover, these vertices generate a tadpole diagram that
contribute to the gluon propagator. Apart from shifting the Gaussian
fixed point of higher order operators in the Yang-Mills sector to an
interacting one, see \cite{Christiansen:2017gtg} for the $U(1)$ case,
it also deforms the gluon contribution to the Yang-Mills
beta function. Its qualitative properties will be discussed later,
as it is important for the large $N_c$ behaviour of the fixed point.
However, a full inclusion is deferred to future work.

\begin{figure}[t]
\includegraphics[width=.9\columnwidth]{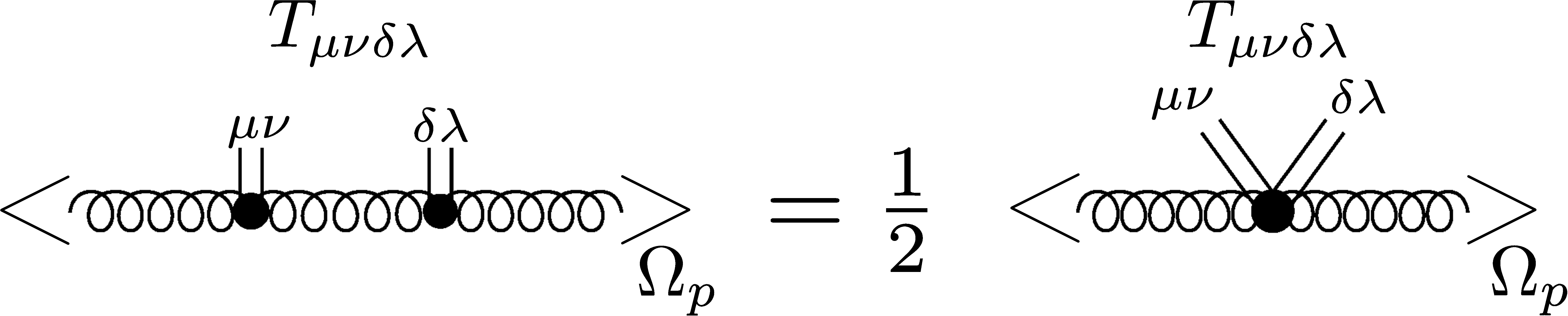}
\caption{Kinematic identity for the one- and two-graviton--two-gluon
  scattering vertices for $r_a=0$ and
  $\Gamma_A^{(2)}\simeq S_A^{(2)}$, taken from 
  \cite{Folkerts:2011jz,FolkertsDiploma}. }
\label{fig:kinID}
\end{figure}

\subsection{Background observables}
\label{sec:background}
The discussion of physics content of background observables and its
relation to gauge- and diffeomorphism invariance has been initiated
for the Yang-Mills--gravity system in
\cite{Folkerts:2011jz,FolkertsDiploma}. There it has been shown that
$\eta_{a,h}= 0$ vanishes for
\begin{align}\label{eq:rArg0}
\0{r_a}{1+r_a} \0{1}{1+r_h}=0\,,
\end{align}
due to a non-trivial kinematic identity. This identity relates angular
averages of one- and two-graviton--two-gluon scattering vertices in the absence
of a gluon regulator $r_a$, see \autoref{fig:kinID}. In other words,
for a combination of regulators that satisfy \eqref{eq:rArg0} the
quantum-gauge and quantum-diffeomorphism symmetry violating effects of
the regulators do not effect the kinematic identity that holds in the
absence of the regulator.

This structure requires some care in the interpretation of the running
of background observables for $k\to\infty$: while the physics
properties of the dynamical fluctuation fields should not depend on
the choice of the regulators, background observables do not
necessarily display physics in this limit. By now we know of many
examples for the latter deficiency ranging from the beta function
of Yang-Mills theory, see \cite{Litim:2002ce}, to the behaviour of the
background couplings in pure gravity,
\cite{Christiansen:2012rx,Christiansen:2014raa,Christiansen:2015rva,
  Christiansen:2016sjn,Denz:2016qks} and matter-gravity systems
\cite{Meibohm:2015twa,Eichhorn:2016esv}. Moreover, we have already
argued that the relation between the dynamical and
the background minimal coupling only holds without modifications for
sufficiently large momenta.

\begin{figure}[t!]
\centering
\includegraphics[width=.71\linewidth]{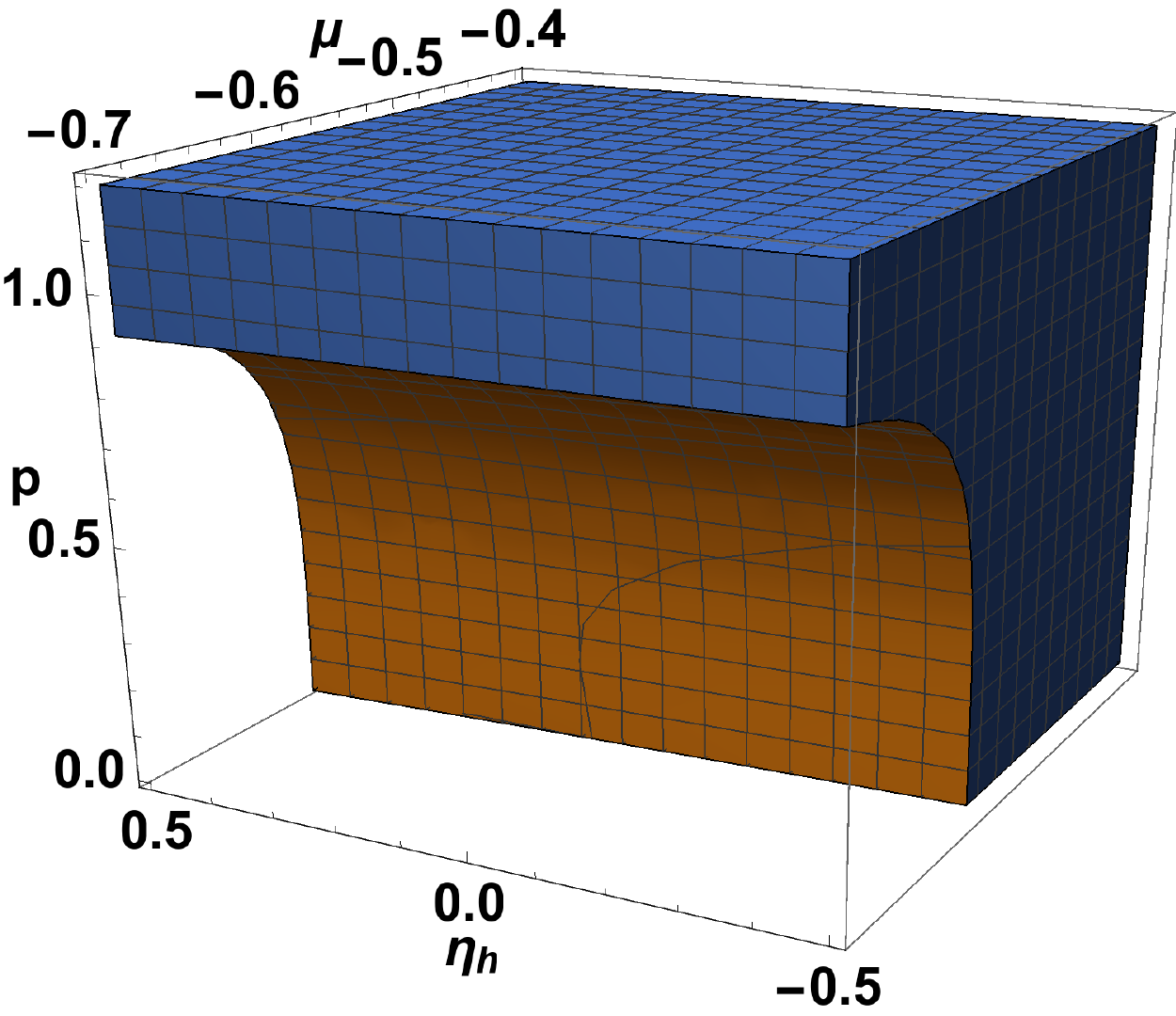}
\caption{ Sign of the graviton contributions to the gluon anomalous
  dimension $\eta_{a,h}$ as a function of
  $\eta_h$, $\mu$, and $p$. The coloured
  region indicates $\mathrm{sgn}\, \eta_{a,h}<0$. At $p=k$ the
  whole displayed region supports asymptotic freedom.}
 \label{fig:Full_Mom_Dep3D}
\end{figure}

In summary, this implies the following for the interpretation of
background observables: we either choose pairs of regulators that
satisfy \eqref{eq:rArg0} or we evaluate background observables for
momentum configurations that are not dominantly affected by the
breaking of quantum-gauge and quantum-diffeomorphism invariance. Here,
we will pursue the latter option that gives us more freedom in the
choice of regulators. For the computation of the graviton contribution
to the running of the Yang-Mills background coupling, this implies that
we have to evaluate the flow of the two-point function for
sufficiently large external momenta,
\begin{align}\label{eq:pKinID}
p^2 \gtrsim k^2\,.  
\end{align}
For these momenta, the three-point function diagrams effectively
satisfy \eqref{eq:rArg0}, and the anomalous dimension $\eta_{a,h}(p^2)$
carries the information about the graviton contribution of the
beta function of the background coupling.

\begin{figure}[t!]
\includegraphics[width=\linewidth]{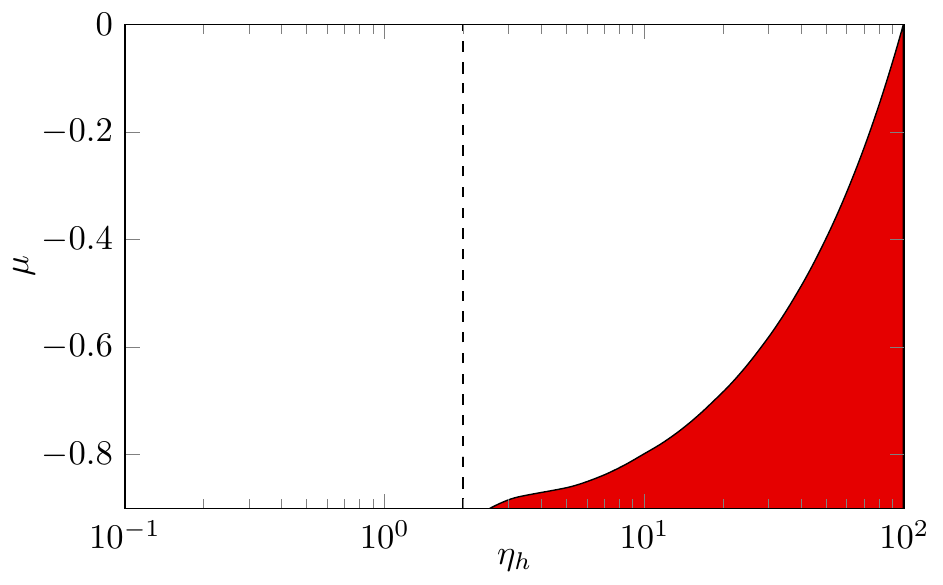}
\caption{Sign of the graviton contributions to the gluon anomalous
  dimension $\eta_{a,h}(k^2)$ as a function of
  $\eta_h$ and $\mu$.  The red region indicates
  $\mathrm{sgn}\,\eta_{a,h}(k^2)>0$ and the loss of asymptotic
  freedom.  The dashed line marks $\eta_h=2$.}
 \label{fig:sign-change-main}
\end{figure}

\subsection{Gravity supports asymptotic freedom}
\label{sec:gravity-to-YM}
The results of the discussion on background observables in the
previous \autoref{sec:background} allow us to access the question of
asymptotic freedom of the minimal Yang-Mills coupling.  With the
construction of the effective action \eqref{eq:trunc_YM_grav}, we obtain a
flow equation for $\partial_t \Gamma^{(aa)}$, which is projected 
with the transverse projection operator
\begin{align}
P_{\mathrm{T}}^{\mu \nu}(p) = \delta^{\mu \nu}- \0{p^\mu p^\nu}{p^2} \,. 
\end{align}
The graviton-induced contributions to the resulting flow equation take
the form
\begin{align}\label{eq:RHS_YM}
  \notag   P_{\mathrm{T}}^{\mu \nu}(p) \, &\partial_t \Gamma_{\mu \nu}^{(aa)}(p) 
                                            =\mathrm{Flow}_h^{(aa)}(p^2) = 
  \\ \notag & Z_a(p^2) \,g \int_q \Big(
              (\dot{r}(q^2)-\eta_{a}(q^2) r(q^2) ) f_a (q,p,\mu)  
  \\ &\quad  + (\dot{r}(q^2)-\eta_{h}(q^2) r(q^2) ) f_h (q,p,\mu) \Big) \, ,
\end{align}
where the terms on the right-hand side originate from diagrams with a regulator insertion in the
gluon and graviton propagator, respectively. The left-hand side is
simply given by
\begin{align}\label{eq:LHS_YM}
P_{\mathrm{T}}^{\mu \nu} \, \partial_t \Gamma_{\mu \nu}^{(aa)}(p) = p^2 
\partial_t Z_a(p^2) \,.
\end{align}
Dividing by $Z_a(p^2)$, one obtains an inhomogeneous Fredholm integral equation of the second
kind for the gluon anomalous dimension,
\begin{align}
\eta_a(p^2) = f(p^2) + g \int\! \0{\mathrm d^4q}{(2 \pi)^4}
\,K\!\left(p,q,\mu,\eta_h \right) \eta_a(q^2) \,.
\label{eq:Fredholm_equ}
\end{align}
This integral equation can be solved using the resolvent formalism by
means of a Liouville-Neumann series. In this work we approximate the
full momentum dependence by evaluating the anomalous dimension in the
integrand in \eqref{eq:Fredholm_equ} at $q^2=k^2$.  This is justified
since the integrand is peaked at $q\approx k$ due to the regulator.
With this approximation, \eqref{eq:Fredholm_equ} can be evaluated
numerically for all momenta.  This approximation was already used in
\cite{Meibohm:2015twa} and lead to results in good qualitative 
agreement with the full momentum dependence.
Details of the full solution are
discussed in App.~\ref{app:integral_equ}.
With the approximation to \eqref{eq:Fredholm_equ}, we investigate
the sign of the graviton contributions to the
gluon propagator.  These
contributions are functions of the gravity couplings,
which in turn depend on the truncation. It is therefore
interesting to evaluate $\eta_{a,h}$ with a parametric dependence on
the gravity couplings, in order to obtain general conditions under
which asymptotic freedom is guaranteed.

The gluon anomalous dimension is of the form
$\eta_a(p^2,g,\mu, \eta_h)$.  In order to avoid the unphysical
regulator dependence potentially induced by the violation of the
kinematical identity \eqref{eq:rArg0} we choose the momentum $p^2 =k^2$ in order to
satisfy \eqref{eq:pKinID}. In summary, this provides us with a minimal
coupling $\alpha_s$,
\begin{align}\label{YM_beta_mom_dep}
\partial_t \alpha_s= \beta_{\alpha_s}= \eta_a(k^2)\,\alpha_s\, .  
\end{align}
As a main result in the present section, we conclude that
\begin{align}\label{eq:YM_beta=AF}
  \beta_{\alpha_s}\leq 0 \qquad  \text{for}\quad  \mu>-1\quad \text{and} \quad  
  \eta_h(k^2)\leq 2 \,. 
\end{align}
The restriction to $\eta_h\leq2$ is also the bound on the anomalous
dimension advocated in \cite{Meibohm:2015twa}. To be more precise,
$\eta_h>2$ only changes the sign of the Yang-Mills beta function in
the limit $\mu\to-1$. For other values of $\mu$, very large values of
$\eta_h$ are necessary in order to destroy asymptotic freedom,
e.g.~for $\mu=-0.4$ the bound is $\eta_h\approx50$ . The precise bound is
displayed in \autoref{fig:sign-change-main}, where the red region indicates
$\beta_{\alpha_s}> 0$.

Despite the necessary restriction to momenta $p^2\gtrsim k^2$ for its
relation to the physical background coupling, we have also evaluated
$\eta_{a,h}$ for more general momentum configurations and a range of
gravity parameters $\mu$ and $\eta_h$: in \autoref{fig:Full_Mom_Dep3D},
the sign of the graviton-induced part of the gluon anomalous
dimension $\eta_{a,h}$ is plotted in the momentum range
$0\leq p^2 \leq k^2$.  For small momenta, $\eta_{a,h}$
changes sign for $\mu\to -1$. Again it can be shown that this does not
happen for regulators with \eqref{eq:rArg0}.

\begin{figure}[t!]
\includegraphics[width=.7\linewidth]{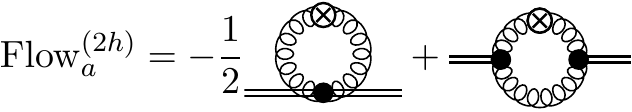}
\caption{Diagrammatic depiction of the gluon contributions to the flow of
the graviton propagator. Wiggly and double lines represent gluon and 
graviton propagators, respectively.}
\label{fig:flow_YM_to_grav_prop}
\end{figure}

In order to understand the patterns behind \autoref{fig:Full_Mom_Dep3D}
and \autoref{fig:sign-change-main}
it is illuminating to examine $\eta_{a,h}(p^2=0)$ for
flat regulators \eqref{eq:flat} with a $p^2$ derivative. It reads
\begin{align}\label{eq:etaAhflatreg}
\eta_{a,h}= -  \0{g}{8 \pi} \left(
\0{8-\eta_a}{1+\mu} -  
\0{4-\eta_h}{(1+\mu)^2} \right) \,. 
\end{align}
The first term on the right-hand side stems from $\partial_t R_{k,a}$
and is positive for $\eta_a< 8$. The second stems from
$\partial_t R_{h,k}$. It is non-vanishing for $\eta_h=0$ and hence
already contributes at one-loop order. Its very presence reflects the
breaking of the non-trivial kinematical identity depicted in
\autoref{fig:kinID} as it is proportional to it. The interpretation of
$\eta_{a,h}$ as the graviton-induced running of the Yang-Mills
background coupling crucially hinges on physical quantum gauge
invariance: it is important to realise that only with the relation
between the auxiliary background gauge invariance and quantum gauge
invariance the latter carries physics. In turn, in the momentum regime
where the kinematical identity is violated, physical gauge invariance
is not guaranteed, and background gauge invariance reduces to an
auxiliary symmetry with no physical content.  Accordingly, one either
has to evaluate $\eta_{a,h}(p^2)$ for sufficiently large momenta
$p^2 \gtrsim k^2$ or utilises regulators that keep the kinematical
identity \autoref{fig:kinID} at least approximately for all momenta.

In summary, \autoref{fig:Full_Mom_Dep3D} and \autoref{fig:sign-change-main} entail that
$\mathrm{sgn}(\eta_{a,h}) < 0$ holds for physically relevant momenta and values of the gravity couplings.
Thus asymptotic freedom is preserved. We have argued that
\eqref{YM_beta_mom_dep} provides the correct definition for the
beta function of the minimal coupling of Yang-Mills
theory with ${\rm sgn} (\beta_{\alpha_s})\leq 0$.  Hence we conclude
that an ultraviolet fixed point in the spirit of the asymptotic safety
scenario is compatible with asymptotic freedom of the minimal coupling
in Yang-Mills theories. In App.~\ref{app:sign-change}, we utilise different approximations
to the gluon anomalous dimension, and we discus in detail the regimes
where it changes the sign in the parameter space of the gravity
couplings.

\section{Yang-Mills contributions to gravity}
\label{sec:YM_to_QG}
This section is concerned with the impact of gluon fluctuations on
the gravity sector. The
fully coupled system is analysed subsequently in \autoref{sec:ASYMG}.

\begin{figure}[t!]
\includegraphics[width=.9\linewidth]{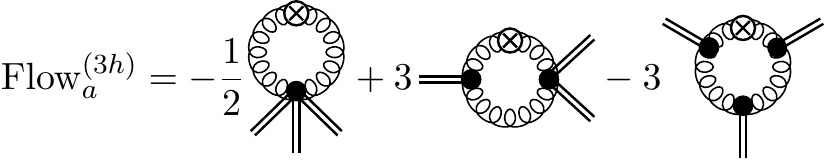}
\caption{Diagrammatic depiction of the gluon contributions to the flow of
the graviton three-point function. 
Wiggly and double lines represent gluon and graviton propagators, respectively.}
\label{fig:flow_YM_to_grav_three}
\end{figure}

\subsection{General structure}
For the question of asymptotic safety, we have 
to investigate the gluon contributions to the graviton
propagator as well as to the graviton three-point function.
This allows us to compute the corrections to the running of 
the gravity couplings $(\mu,g,\lambda_3)$ due to gluon fluctuations.  

The gluon corrections to the graviton two- and three-point
function split analogously to the graviton corrections to Yang-Mills
theory in the preceding section, since for any graviton $n$-point function the
structure is given by
\begin{align}
\mathrm{Flow}^{(nh)} =  \mathrm{Flow}^{(nh)}_h +
\mathrm{Flow}^{(nh)}_a \, ,
\end{align}
with graviton  and gluon contributions denoted by $\mathrm{Flow}^{(nh)}_h$ 
and $ \mathrm{Flow}^{(nh)}_a$, respectively. For example, the
gluon contributions to the flow of the graviton two- and
three-point function are depicted in
\autoref{fig:flow_YM_to_grav_prop} and
\autoref{fig:flow_YM_to_grav_three}.
Accordingly, the beta function for Newton's coupling including gluon
corrections has the structure 
\begin{align}\label{eq:beta_g_YM}
\partial_t g ={}& \left(2 +3 \eta_h\right)g   \\ 
&+ g^2 \Bigl( A_h(\mu,\lambda_3) + \eta_h B_h(\mu,\lambda_3) +
C_a + \eta_a D_a
\Bigr) \,,\notag
\end{align}
where we have used the identifications \eqref{eq:gn}.  
In \eqref{eq:beta_g_YM}, $A_h$ and $B_h$ originate from graviton loops and they
depend on $\mu$ and $\lambda_3$, while
$C_a$ and $D_a$ are generated by gluon
loops and are just numbers. 
Similarly the beta function for $\lambda_3$ has the structure
\begin{align}\label{eq:beta_lam3_YM}
\partial_t \lambda_3 ={}& \left(-1 +\023 \eta_h + \0{\partial_t g}{2g}\right)\lambda_3  \\
&+ g \Bigl( E_h(\mu,\lambda_3) + \eta_h F_h(\mu,\lambda_3) +G_a + \eta_a H_a
\Bigr) \,. \notag
\end{align}
Throughout this chapter we display the anomalous dimensions $\eta_h,\eta_a$ as momentum independent.
Note, however, that they are momentum dependent and we approximate their momentum dependence by
evaluating them at $p=k$ if they appear in an integral, see \cite{Meibohm:2015twa} for details.

\begin{figure*}[t!]
\includegraphics[width=\linewidth]{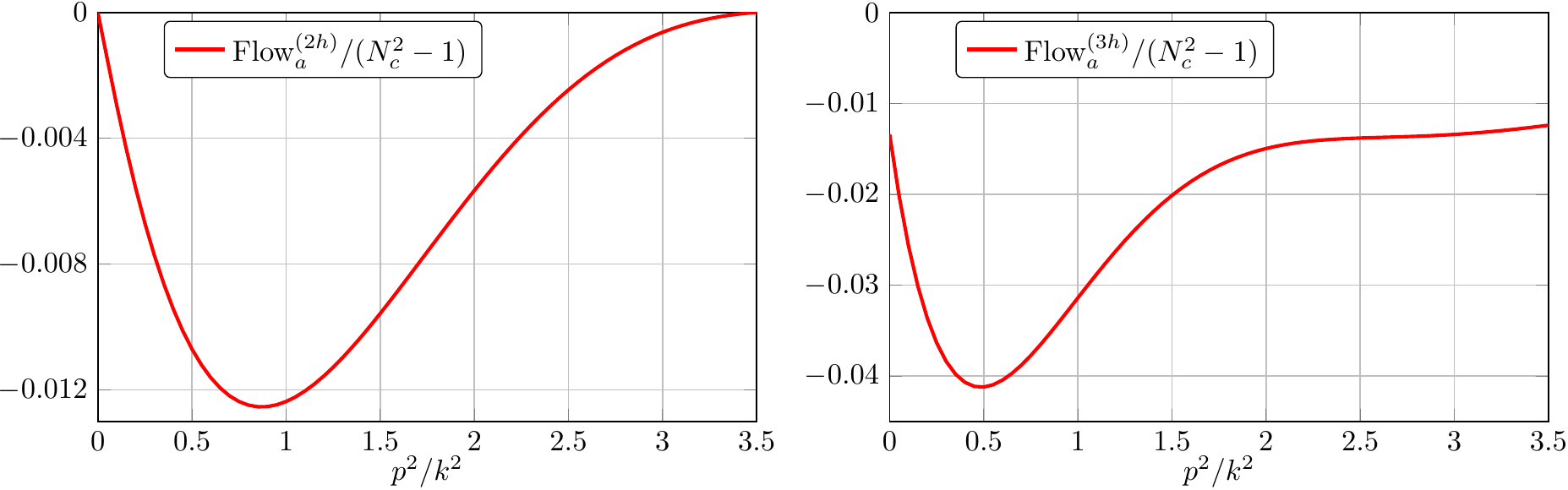}
\caption{The momentum dependence of $\mathrm{Flow}_a^{(2h)}/(N_c^2 -1)$ (left) 
  and $\mathrm{Flow}^{(3h)}_a/(N_c^2 -1)$ (right) for $g=1$ and
  $\eta_a =0$ on the right-hand side of the flow.}
\label{fig:MomDp_YM_to_grav}
\end{figure*}

Moreover, the Yang-Mills
contributions to the graviton propagator enter the above beta function
\eqref{eq:beta_g_YM} via the graviton anomalous dimension $\eta_h$ and
the graviton mass parameter $\mu$. These equations
have the general form
\begin{align}\label{eq:beta_mu_YM} 
\eta_h ={}& g\Bigl( I_h(\mu,\lambda_3)+ \eta_h J_h(\mu,\lambda_3) +K_a +\eta_a L_a\Bigr) \,, \notag \\
\partial_t \mu ={}& (\eta_h -2) \mu 
\\
&\,
+g\Bigl(M_h(\mu,\lambda_3) + N_h(\mu,\lambda_3)\eta_h + O_a+ \eta_a P_a\Bigr) \, , \notag
\end{align}
where again all pure gravity contributions are labelled with an index $h$ and
the one generated by gluons with an index $a$.
Note again that all the Yang-Mills contributions do not depend on $\mu$ and $\lambda_3$, as the
corresponding diagrams do not
involve graviton propagators and pure graviton vertices, see \autoref{fig:flow_YM_to_grav_prop} and
\autoref{fig:flow_YM_to_grav_three}. In particular, this implies that these
terms have no $1/(1+\mu)$ singularity in the limit $\mu \to -1$.
Furthermore, all these diagrams contain a
closed gluon loop, and hence, all the factors in the above equations with an
index $a$ are proportional to $N^2_c -1$.

\subsection{Contributions to the graviton propagator}
\label{sec:ym-to-prop}
The gluon contribution to the graviton propagator has been
studied in a derivative expansion around $p^2=0$ in
\cite{FolkertsDiploma} where it was shown that this
projection is insufficient due to the non-trivial momentum dependence
of the flow. The latter is characterized by a dip at $p^2 \approx k^2$. It
has been shown in \cite{Christiansen:2012rx} that this structure is
also present in the full flow, i.e.\ including the graviton
contributions and that projections at momentum scales close to the
cutoff are necessary, see also
\cite{Christiansen:2015rva,Denz:2016qks}. We have rederived the
momentum dependence of $ \mathrm{Flow}_a^{(2h)}(p^2)$, 
see \autoref{fig:MomDp_YM_to_grav}.

For the projection at $p^2=0$ and flat regulators \eqref{eq:flat}, we
rederive the result of \cite{FolkertsDiploma} and obtain for the
momentum-independent part
\begin{align}\label{eq:cha}
\mathrm{Flow}_a^{(2h)}\left(p^2=0\right) =  g Z_h(N_c^2-1)\0{1}{60 \pi}
\eta_a \, . 
\end{align}
Surprisingly, this contribution is proportional to $\eta_a$.  This
happens due to a cancellation between both diagrams displayed in
\autoref{fig:flow_YM_to_grav_prop}.  Note that this cancellation only
occurs for the flat regulator.  For other regulators the contribution
can be either positive or negative.  This is discussed in 
App.~\ref{app:ym-sign-to-dot-mu} and will play a crucial r\^ole in the
later analysis.

For the computation of the graviton anomalous dimension, we resort 
to a finite difference projection, which is 
of the general form
\begin{align}
\0{\mathrm{Flow}_a^{(2h)}({p^2_1})-\mathrm{Flow}^{(2h)}_a({p^2_2})}{p^2_1-p^2_2}= g Z_h (N_c^2-1)(\alpha + \beta\,\eta_a ) \,, 
\end{align}
where $\alpha$ and $\beta$ depend only on $p_1$ and $p_2$. This is
rooted in the fact that there are only internal gluon propagators and
graviton-gluon vertices, and these do not depend on $\lambda_3$ and $\mu$ 
as discussed in the last section. For
$p_2=0$ and $p_1 \rightarrow p_2$, i.e.\ a $p^2$-derivative at $p^2=0$, we
obtain
\begin{align}\label{eq:0}
\alpha = \beta =  - \0{1}{12 \pi}  \approx -0.027 \,.
\end{align}
For a finite difference with $p^2_1=k^2$ and $p_2=0$,
we obtain
\begin{align}\label{eq:k}
\alpha  \approx-0.012\,, \qqquad  \beta  \approx -0.0033\,.
\end{align}
\eqref{eq:0} and \eqref{eq:k} display the gluon contribution to $-\eta_h$; thus,
the gluon contribution to $\eta_h$ is positive independent of the momentum projection
scheme. Note however that \eqref{eq:0} and \eqref{eq:k} display a qualitatively different
behaviour, and \eqref{eq:k} is the correct choice due to the
momentum dependence of the flow. This has already been observed in the
pure gravity computations in
\cite{Christiansen:2012rx,Christiansen:2014raa,Christiansen:2015rva,Denz:2016qks}
and emphasises the importance of the momentum-dependence.
In this work we use a finite difference between $p^2_1=p^2$ and $p^2_2=-\mu k^2$
for the equation of $\eta_h(p^2)$, see \cite{Christiansen:2014raa,Meibohm:2015twa} for details.

\subsection{Contributions to the three-point function}
\label{sec:YM_to_grav_three}
The contributions to the graviton three-point function enter
the beta function of the Newton's coupling
$g$ \eqref{eq:beta_g_YM} via $C_a$ and $D_a$ and
the beta function of $\lambda_3$ \eqref{eq:beta_lam3_YM} via $G_a$ and
$H_a$.  The diagrammatic representation of these contributions is
shown in \autoref{fig:flow_YM_to_grav_three}.  Here, the contribution
to $\partial_t g$ is the momentum dependent part and the contribution
to $\partial_t \lambda_3$ in the momentum independent part to the
graviton three-point function.  For the projection on the couplings
$g$ and $\lambda_3$, we use precisely the same projection operators as in
\cite{Christiansen:2015rva}.  These are different projection operators for $g$
and $\lambda_3$, and we mark this with an index $G$ and $\Lambda$ in
the following.

We have seen in the previous 
sections, that the momentum dependence of the flow plays a crucial r\^ole, and 
key properties may be spoiled if non-trivial momentum dependence is not taken 
into account properly. Therefore, we resolve the momentum dependence 
of the contributions $\mathrm{Flow}_{G,a}^{(3h)}(p^2)$, 
which is shown in the right panel of \autoref{fig:MomDp_YM_to_grav}.
Interestingly, the contribution is peaked at $p^2=\012k^2$
and is not well described by $p^2$ in the region 
$0\leq p^2\leq k^2$. 
Because of this non-trivial structure, the contribution to $\partial_t g$ depends
on the momenta where it is evaluated.
For general momenta $p^2_1$ and $p^2_2$, we obtain
\begin{align}\label{eq:YM_contr_explcitly}
  \0{\mathrm{Flow}^{(3h)}_{G,a}(p_1^2)- \mathrm{Flow}^{(3h)}_{G,a}(p_2^2)}{p_1^2-p_2^2}
  = g^{\032}
  Z_h^{\032} (N_c^2-1)( \gamma + \delta\, 
  \eta_a ) \, , 
\end{align}
where $\gamma$ and $\delta$ again only depend on $p^2_1$ and $p^2_2$.
Evaluated as derivatives, i.e., $p_2^2=0$ and $p_1^2\to 0$, we arrive at 
 \begin{align}\label{eq:gadel0}
\gamma  &= -\07{30\pi}\approx -0.074\,, 
&
\delta  &= -\01{570\pi}\approx-0.00056\,.
\end{align}
With $p_1^2=k^2$ and $p_2^2=0$, they are given by  
\begin{align}\label{eq:gadel}
\gamma  &\approx-0.018\,, 
&  \delta  &\approx -0.0014\,.
\end{align}
As in the case of the gluon propagator, the sign of the derivative
definition agrees with the bi-local one but they differ strongly in
their magnitude.  In the present work, we use \eqref{eq:gadel}.
The contribution to $\lambda_3$ is always evaluated at vanishing momentum.
We obtain
\begin{align}
 \mathrm{Flow}_{\Lambda,a}^{(3h)}\left(p^2=0\right) = 
 g^{\032} Z_h^{\032} (N_c^2-1) \0{ 3 - \eta_a }{60 \pi} \,.
 \label{eq:ym-to-lam3}
\end{align}

\subsection{Mixed graviton-gluon coupling}
So far, we have only considered pure gluon and pure graviton
correlation functions in the coupled Yang-Mills--gravity system.
Indeed, the results that will be presented in \autoref{sec:ASYMG} are
based on precisely these correlation functions, and other couplings are
identified according to \eqref{eq:gn}.  In \autoref{sec:general}, we will
then discuss the stability of the results under extensions of the
truncation.  In particular, we will have a look at the inclusion of a
flow equation for the graviton--two-gluon coupling $g_a$.

The flow equation for $g_a$ is derived analogously to the $g_3$
coupling from three-graviton vertex: we build the projection operator
from the classical tensor structure $S^{(haa)}$ with a transverse
traceless graviton and two transverse gluons.  This projection
operator is contracted with both sides of the flow equation for
this specific vertex.  The equation is further evaluated at the
momentum symmetric point \cite{Christiansen:2015rva}.  The resulting
$p^2$ part gives the flow equation for $g_a$.  We obtain an analytic
flow equation for $g_a$ by a $p^2$ derivative at $p^2=0$.  The
resulting flow equation is given in App.~\ref{app:floweq}.

For the computations in \autoref{sec:general}, we use the preferred
method of finite differences.  In particular, we choose the evaluation
points $p^2=k^2$ and $p^2=0$.  With this method, we do not obtain
analytic flows but we take more non-trivial momentum dependences into
account \cite{Christiansen:2015rva,Denz:2016qks}.  The computation is
simplified by the fact that the present flow is actually vanishing at
$p^2=0$.  Consequently, the finite difference equals to an evaluation
at $p^2=k^2$, and the momentum derivative gives the same result as a
$1/p^2$ division.  

\begin{figure*}[t!]
\includegraphics[width=\textwidth]{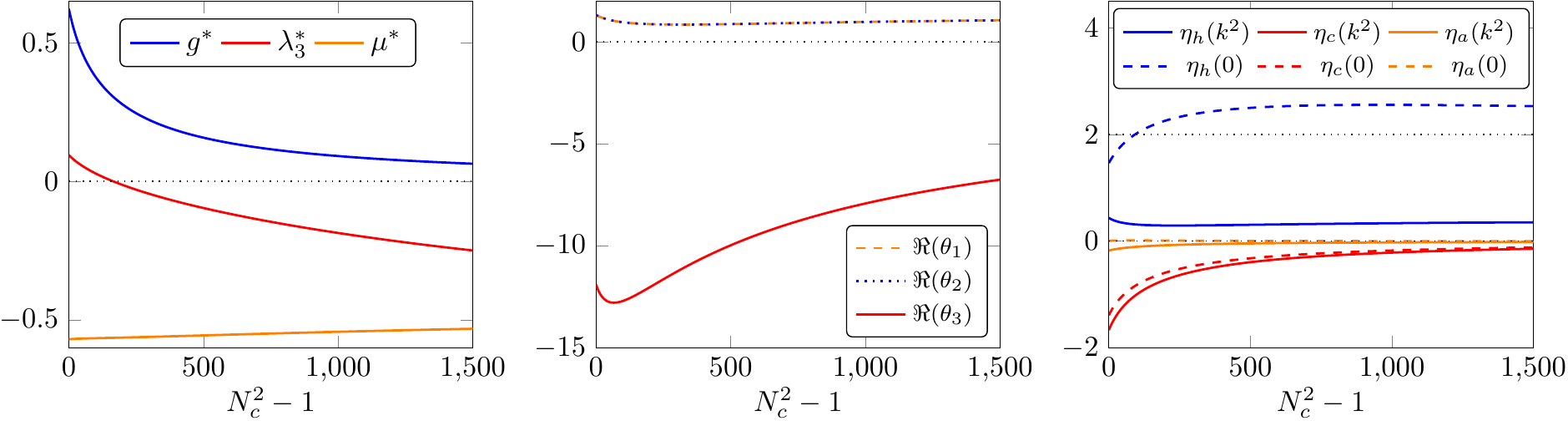}
\caption{Properties of the UV fixed point as a function of $N_c^2-1$
  in the uniform approximation with one Newton's coupling.
  Displayed are the fixed point values (left panel), the critical
  exponents (central panel), and the anomalous dimensions (right
  panel).}
\label{fig:FPs-uniform}
\end{figure*}

\subsection{Momentum locality}
We close this section with a remark on the momentum locality
introduced in \cite{Christiansen:2015rva} as a necessary condition for
well-defined RG flows. It was shown to be related
to diffeomorphism invariance of the theory. It entails that flows
should not change the leading order of the large momentum behaviour of
correlation functions. 

The asymptotics of the diagrams for the graviton two-point function, 
ordered as displayed in \autoref{fig:flow_YM_to_grav_prop}, are
\begin{align}\nonumber 
\text{Diag}_{\,1}^{(2h)}(p^2\to\infty) &=  - g \0{8-\eta_a}{12 \pi} \,,\\
\text{Diag}_{\,2}^{(2h)}(p^2\to\infty) &=  g \0{8-\eta_a}{12 \pi} \,,
\end{align}
while the asymptotics for the graviton three-point function,
again ordered as displayed in \autoref{fig:flow_YM_to_grav_three},  are
\begin{align}\nonumber 
\text{Diag}_{\,1}^{(3h)}(p^2\to\infty) &=  -g^{3/2} \0{8-\eta_a}{19 \pi}\,,\\\nonumber 
\text{Diag}_{\,2}^{(3h)}(p^2\to\infty) &=  g^{3/2} \0{4 (8-\eta_a)}{19 \pi} \,,\\
\text{Diag}_{\,3}^{(3h)}(p^2\to\infty) &=  - g^{3/2} \0{3 (8-\eta_a)}{19 \pi} \,.
\end{align}
Consequently we again have a highly non-trivial cancellation between
different diagrams, which leads to the property of momentum locality.
In summary, we assert
\begin{align}
\lim_{p^2/k^2\rightarrow \infty} \0{\partial_t
\Gamma^{(2h,3h)}(p^2)}{\Gamma^{(2h,3h)} (p^2) } 
=0 \, ,
\end{align} 
at the symmetric point in the transverse traceless mode.
Hence, the full flows of the graviton two- and
three-point functions including Yang-Mills corrections are
momentum local.

\section{Asymptotic safety of Yang-Mills--gravity}
\label{sec:ASYMG}
In this section, we provide a full analysis of the ultraviolet fixed
point of the coupled Yang-Mills--gravity system. It is characterised by
the non-trivial fixed point of Newton's coupling $g$, the coupling
of the momentum-independent part of the graviton three-point function
$\lambda_3$, and the graviton mass parameter $\mu$ while the minimal
gauge coupling vanishes, $\alpha_s=0$.

\subsection{Finite \texorpdfstring{$N_c$}{Nc}}
The fully coupled fixed point shows some remarkable features.  The
fixed point values are displayed in the left panel of
\autoref{fig:FPs-uniform}.  The fixed point value of the graviton
mass parameter remains almost a constant as a function of $N_c$.  The
Newton's coupling is approaching zero, while $\lambda_3^*$ becomes
slowly smaller and crosses zero at $N_c^2\approx 166$.  This
behaviour can be understood from the equations:
the leading contribution from Yang-Mills to $\partial_t \mu$ cancels
out, and only a term proportional to $\eta_a$ remains, see \eqref{eq:cha}.
The latter is small at the fixed point, and hence, the effect on
$\partial_t \mu$ is strongly suppressed.  The fall off of $g^*$ and
$\lambda_3^*$ is explained by the respective contribution in
the flow equations, see \eqref{eq:gadel} and \eqref{eq:ym-to-lam3}.

The critical exponents of the fixed point, 
which are given by minus the eigenvalues of the stability matrix,
are displayed in the central
panel of \autoref{fig:FPs-uniform}.  They remain stable over the
whole investigated range.  Two critical exponents form a complex
conjugated pair.  The real part of this pair is positive and thus
corresponds to two UV attractive directions.  The third critical exponent
is real and negative and corresponds to a UV repulsive direction.
The eigenvector belonging to the latter exponent points approximately
in the direction of $\lambda_3$, which is in accordance with pure 
gravity results \cite{Christiansen:2015rva}.

In the right panel of \autoref{fig:FPs-uniform}, we show the
anomalous dimensions at the fixed point, evaluated at $p^2=0$ and
$p^2=k^2$.  The ghost and gluon anomalous dimensions tend towards zero for increasing $N_c$.  
Most importantly,
$\eta_a(k^2)$ is always negative, which is a necessary condition for
asymptotic freedom in the Yang-Mills sector.  The graviton anomalous
dimension does not tend towards zero.  At $p^2=k^2$, it is getting
smaller with an increasing $N_c$ despite the positive gluon
contribution \eqref{eq:k}.  The reason is that the anomalous
dimension is also proportional to $g^*$, which is decreasing, and
this effect dominates over the gluon contribution.  At $p^2=0$, on the other hand,
the gluon contribution is also positive but larger in value, see
\eqref{eq:0}, and consequently, dominates over the decrease in $g^*$.
$\eta_h(0)$ is increasing, crosses the value 2 and starts to 
decrease again for large $N_c$.
As mentioned in \eqref{eq:nonReg}, $\eta < 2$ is a bound on regulators
that are proportional to the respective wave function renormalisation.
In our case, $\eta_h(0)$ exceeds the value 2
just slightly and remains far from the strict bound, which is
$\eta_h<4$, see \cite{Meibohm:2015twa} for details. 

\begin{figure}[t!]
\includegraphics[width=.9\linewidth]{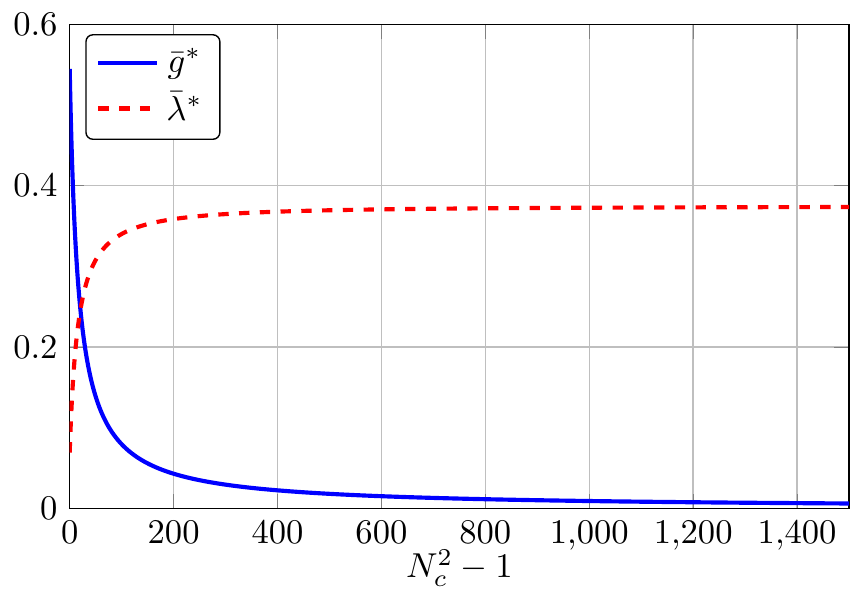}
\caption{Displayed are the background couplings $\bar g^*$ and
  $\bar \lambda^*$ as a function of $N_c^2-1$ evaluated at the UV fixed
  point displayed in \autoref{fig:FPs-uniform}.  The coupling
  $\bar g^*$ is going to zero with $\01{N_c^2}$ and $\bar \lambda^*$ goes to
  the constant $0.38$, see \eqref{eq:Nclimitev}.  }
\label{fig:FP-Background}
\end{figure}

The fixed point values of the background couplings are displayed in
\autoref{fig:FP-Background}.  The equations for the pure gravity part
are identical to the ones in \cite{Denz:2016qks} and the gluon part is
identical to the one in \cite{Dona:2012am}.  In this setting, the
background couplings behave very similar to the dynamical ones.  The
background Newton's coupling goes to zero with $1/N_c^2$ while the
background cosmological constant goes to a constant for large $N_c$.
Interestingly, the background coupling approach their asymptotic behaviour
faster than the dynamical ones.

\subsection{Large \texorpdfstring{$N_c$}{Nc} scaling}
\label{sec:largeNc}
\begin{figure}[t!]
\includegraphics[width=.9\linewidth]{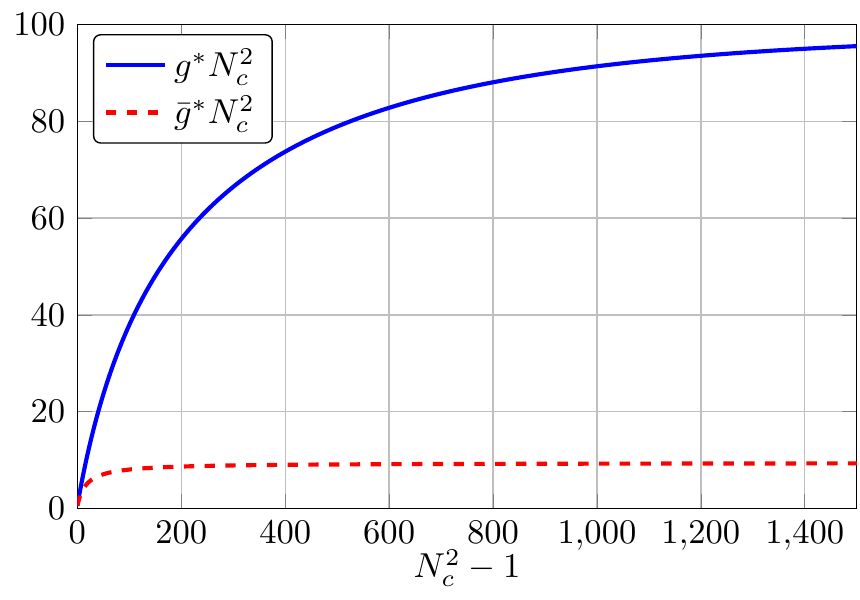}
\caption{Displayed are the fixed point 't Hooft couplings $g^* N_c^2$ and
  $\bar g^* N_c^2$ as a function of $N_c^2-1$.  The couplings approach the
  asymptotic values $g^* N_c^2 \to 89$ and $\bar g^* N_c^2 \to 9.4$, see
  \eqref{eq:NclimitFP} and \eqref{eq:Nclimitev}.}
\label{fig:T-Hooft-coupling}
\end{figure}

In the limit $N_c \to \infty$, the couplings approach the fixed point
values
\begin{align}
g^* &\to \0{89}{N_c^2} + \0{8.0 \cdot 10^4}{N_c^4} \,,
&
\mu^* &\to -0.45 - \0{3.3 \cdot 10^2}{N_c^2}\,, \notag\\
\lambda_3^* &\to -0.71 + \0{2.4 \cdot 10^3}{N_c^2} \,.
\label{eq:NclimitFP}
\end{align}
As expected, the 't Hooft coupling $g^* N_c^2$ is going to a constant
in the large $N_c$ limit.  This behaviour is also displayed in
\autoref{fig:T-Hooft-coupling} for finite $N_c$.  Remarkably, $\mu^*$
and $\lambda_3^*$ remain finite.  In the $\lambda_3$ equation, this
originates from a balancing of the gluon contribution with the
canonical term.  In the $\mu$ equation, on the other hand, all
contributions go to zero in leading order and the fixed point value of
$\mu$ follows from the second order contributions.  The asymptotic
anomalous dimensions follow as
\begin{align}
 \eta_h (0) &\to 2 + \0{2.7\cdot 10^3 }{N_c^2} \,,
 &
 \eta_h(k^2) &\to  0.36 + \0{2.9 \cdot 10^2 }{N_c^2} \,,\notag\\
 \eta_c(0) &\to -\0{1.3 \cdot 10^2}{N_c^2} \,,
 &
 \eta_c(k^2) &\to -\0{1.5 \cdot 10^2}{N_c^2} \,, \notag\\
 \eta_a(0) &\to -\0{8.7}{N_c^2}  \,,
 &
 \eta_a(k^2) &\to -\0{22}{N_c^2}  \,,
 \label{eq:Nclimiteta}
\end{align}
which satisfy the bounds $\eta_i\leq 2$ necessary for the
consistency of the regulators that are proportional to
$Z_h, Z_c, Z_a$.  Note that only the graviton anomalous dimension is
non-vanishing in this limit.  Importantly, the gluon anomalous
dimension approaches zero from the negative direction, which means
that it supports asymptotic freedom in the Yang-Mills sector.  The
asymptotic value $\eta_h(0)=2$ follows directly from the demand that
all contributions in the $\mu$ equation have to go to zero in leading
order, as discussed in the last paragraph.
The critical exponents are given by
\begin{align}
  \theta_{1,2} &\to 1.2 \pm 2.1 i + \0{(1.1 \mp 5.6 i)
                 \cdot 10^3}{N_c^2} \,,\notag\\
  \theta_{3} &\to -2.3 - \0{14\cdot 10^3}{N_c^2} \,. 
  \label{eq:Nclimitev}
\end{align}
The fixed point has two attractive and one repulsive direction for all
colours. Remarkably, the values of the critical exponents remain of order
one. The background couplings approach the values
\begin{align}
 \bar g^* &\to \0{9.4}{N_c^2} -\0{1.3 \cdot 10^2}{N_c^4} \,, 
 \qquad
 \bar \lambda^* \to 0.38  - \0{1.4}{N_c^2} \,.
 \label{eq:Nclimitbackgr}
\end{align}
Again, the background 't Hooft coupling $\bar g^*N_c^2$ remains finite
in the large $N_c$ limit, which is also displayed in
\autoref{fig:T-Hooft-coupling}.

In summary, we have found a stable UV fixed point with two attractive directions.
The fixed point values, the critical exponents and the anomalous dimensions are of order one.
In \autoref{fig:FPs-uniform} we display this behaviour up to $N_c^2=1500$,
and in this section, we have augmented this with a solution for $N_c\to \infty$.
Consequently, we conclude that the system is asymptotically safe in the gravity sector 
and asymptotically free in the Yang-Mills sector for all $N_c$. 


\subsection{Decoupling of gravity-induced gluon self-interactions}
\label{sec:grav-induced}
It has been advocated in \cite{Eichhorn:2012va} that interacting
matter-gravity systems necessarily contain self-interacting matter fixed
points. This has been investigated in scalar, fermionic and Yukawa systems in, e.g., 
\cite{Meibohm:2016mkp,Eichhorn:2016esv,Eichhorn:2017eht}. 

\begin{figure}[t!]
\includegraphics[width=.9\linewidth]{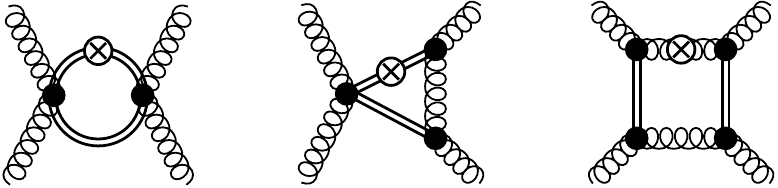}
\caption{Diagrammatic depiction of the graviton induced higher-order gluon interactions.
Wiggly and double lines represent gluon and graviton propagators, respectively.}
\label{fig:higher-ym}
\end{figure}

Recently, also a Yang-Mills--gravity system with an Abelian $U(1)$ gauge
group has been investigated \cite{Christiansen:2017gtg}. It was found
that that the coupling of the fourth power of the field strength,
$F^4$, takes a finite fixed point value, while the minimal coupling
that enters the covariant derivative can be asymptotically free. As
already mentioned before in \autoref{sec:QG_to_YM}, the same
happens in Yang-Mills--gravity systems. In particular, we are led to
\begin{align}\label{eq:F4}
w_2^{*}\, (\tr F_{\mu\nu}^2)^2 + v^*_4\, \tr F_{\mu\nu}^4 \,,
\end{align}
with $w_2^* \neq 0$ and $v_4^*\neq 0$ without non-trivial
cancellations. A quantitative computations of these fixed point
couplings is deferred to future work. Here, we simply discuss their
qualitative behaviour: even if not present in the theory, the
couplings $w_2$ and $v_4$ are generated by diagrams with the
exchange of two gravitons, see \autoref{fig:higher-ym}. In leading order, these diagrams are
proportional to
\begin{align}\label{eq:w2v4}
\0{g^2 }{(1+\mu)^3}\propto \0{1}{N_c^4}\to 0 \,,
\end{align}
and vanish in the large $N_c$ scaling of \eqref{eq:NclimitFP}. It is
simple to show that the further diagrams in the fixed point equations
of $w_2, v_2$ proportional to $w_2, v_2$ decay even faster when using
\eqref{eq:w2v4} for the diagrams. 

Finally, we get additional gluon tadpole contributions proportional to
$\omega^*_2, v^*_4$ for the running of the Yang-Mills
beta function. In leading order these contributions are
proportional to $N_c^2$ due to a closed gluon loop. Together with the
fixed point scaling of $\omega^*_2, v^*_4$ in \eqref{eq:w2v4} this leads
to a $1/N_c^2$ decay of these contributions. They have the same large
$N_c$ scaling as the pure gravity contributions but also share the
same negative sign supporting asymptotic freedom, see
\cite{Christiansen:2017gtg} for a study in $U(1)$ theories. 

We close this chapter with a qualitative discussion of the stability
for the interacting fixed point: as
$\omega_2,v_2$ do not couple into the pure gravity subsystem, the
stability matrix is skew symmetric, and the eigenvalues are computed
in the respective sub-systems. Both, the gravity as well as the
$\omega_2,v_4$ sub-systems are stable in the limit $g\to 0$. 

This concludes our analysis of the large $N_c$ behaviour of quantum
gravity with the flat regulator and the identification \eqref{eq:gn}. As
expected, Newton's coupling $g$ shows the $1/N_c^2$ behaviour
discussed in \autoref{sec:as-af}. 

\begin{figure*}[t!]
\includegraphics{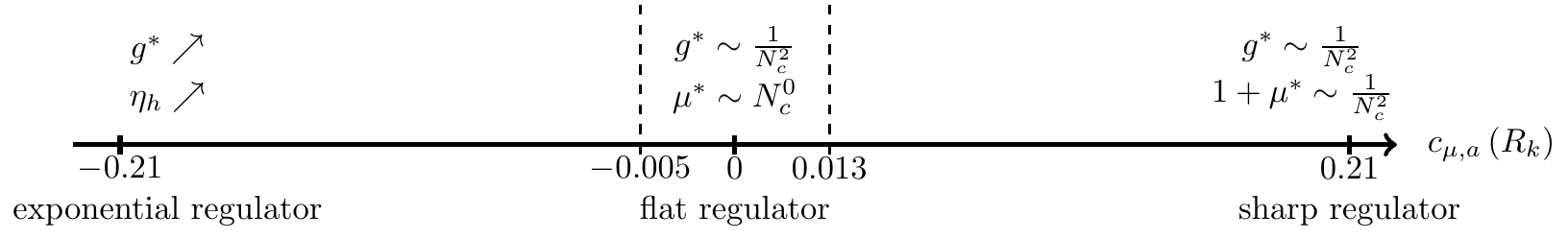}
\caption{Schematic picture of the dynamical scale readjustment 
  mechanisms as a function of the coefficient $c_{\mu,a}(R_k)$.}
\label{fig:regulator-scheme}
\end{figure*}

\section{UV dominance of gravity }
\label{sec:general} 
\subsection{Dynamical scale fixing}
\label{sec:dynscale}
In \autoref{sec:ASYMG}, we used the identifications of all Newton's
couplings \eqref{eq:gn}. In the present chapter, we discuss the general
case without this identification. We provide a comprehensive
summary of results and the underlying structure, more details can be
found in App.~\ref{app:scaling}. While we have argued in
\autoref{sec:as-af} that the present Yang-Mills--gravity system, as well as
all free-matter--gravity systems are asymptotically safe, the
interesting question is how and if at all in the present
approximation this is dynamically observed.

Within the iterative procedure in \autoref{sec:as-af}, we arrived at a
fixed point action that is identical to that of the pure gravity
sector with fixed point values for $g_n^*$, 
$\lambda_n^*$, and $\mu^*$. We also have $g_a = g_3$
due to the expansion of the metric $g_{\mu\nu}=\bar g_{\mu\nu}+\sqrt{g_3}\, k^2 h_{\mu\nu}$ with
$k=k_h$. Note also that in such a two-scale setting with $k_h$ and
$k_a$, the latter rather is to be identified with
$k_a^{\text{\tiny UV}}$ and not with $k_a^{\text{\tiny IR}}$.  As
the effect of the latter has been absorbed in a renormalisation of 
Newton's coupling prior to the integrating out of graviton fluctuations
(or rather their suppression with $k_h\to\infty$), this sets the
graviton cutoff scale $k_h=k$ as the largest scale in the system. This
leads to \eqref{eq:kh-ka} that effectively induces
\begin{align}
\label{eq:kh-ka-rep}
k^2\simeq  N_c^2 k^2_a \,,
\end{align}
in the large $N_c$ limit. Note that with a rescaling of our unique
cutoff scale in \autoref{sec:ASYMG} with $N_c^2$ we already arrive at
the $N_c$-independent fixed-point values \eqref{eq:NclimitFP}. The large
values come from dropping the $N_c$-independent prefactor in the ratio
$G/G_{\text{\tiny eff}}$. The latter fact signals the unphysical
nature of fixed point values, which within this two-scale setting
also extends to the product
$g^* \lambda^*$, typically used in the literature as a potentially
rescaling-invariant observable. 

Despite \eqref{eq:kh-ka} being a natural relative scale setting, without
any approximation the full system of flow equations with $k_h=k_a$
should adjust itself dynamically to this situation with
$g_a^* \sim g_c^* \sim g^*$ and with $g^*\propto 1/N_c^2$
in the large $N_c$ limit. In the present approximation this can happen
via two mechanisms that both elevate the graviton fluctuations to the
same $N_c$ strength as the gluon fluctuations: the graviton propagator
acquires a $N_c$ scaling 
\begin{align}\label{eq:gravNc}
k^2 G_{h}(p^2=0) =\0{1}{Z_{h}}\0{1}{1+\mu}  \propto N_c^2\,, 
\end{align}
after an appropriate rescaling of the couplings, for more details see
App.~\ref{app:scaling}. We proceed by discussing the two 
dynamical options that the system has to generate the $N_c$ scaling in \eqref{eq:gravNc}: 

\begin{itemize}
\item[(1)] Evidently, \eqref{eq:gravNc} can be achieved via
\begin{align}\label{eq:1+mu}
\mu^*\propto  -1 + c_+/N_c^2\,, 
\end{align}
with a positive constant $c_+$. Note that \eqref{eq:1+mu} is not
present in the fixed point results in \autoref{sec:ASYMG}. Accordingly,
adding the fixed-point equation for $g_a$ has to trigger this 
running. Below we shall investigate this possibility in more detail. 
\item[(2)] The $N_c$ scaling can also be stored in $1/Z_{h}$. As we have
chosen regulators that are proportional to $Z_{h}$, this leads to
an effective elimination of $Z_{h}$ from the system; its only
remnant is the anomalous
dimension $\eta_h$ in the cutoff derivative. Since
$1/Z_{h}\propto (k^2)^{\eta_h/2-1}$, the anomalous dimension
$\eta_h$ has to grow large and positive in order to effectively
describe the $N_c$ scaling in \eqref{eq:gravNc}, 
\begin{align}\label{eq:etainf}
  \eta_h\to \infty\,. 
\end{align}
In the present setting with $R_{h,k} \propto Z_{h}$, this option
cannot be investigated as \eqref{eq:etainf} violates the bound
\begin{align}\label{eq:bound}
  R_{h,k} \propto Z_{h}\qquad \Rightarrow\qquad \eta_h<2\,,
\end{align} 
for the regulator. For $\eta_h>2$, the regulators of type
\eqref{eq:bound} cannot be shown to suppress UV degrees of freedom
anymore in the limit $k\to\infty$ as $\lim_{k\to\infty} R_k(p^2)\to 0$
for $\eta_h>2$. This bound was introduced and discussed in
\cite{Meibohm:2015twa} within the scalar-gravity system, where
$\eta_h$ grows beyond this bound for the number of scalars $N_s$
getting large. It was stated there that the stability of the
scalar-gravity system could not be investigated conclusively since the
regulator cannot be trusted anymore. In the light of the present
results and discussion, we know that the free-matter system is asymptotically
safe. Then, the growing $\eta_h$ signals that the system wants
to accommodate \eqref{eq:gravNc} with a growing $1/Z_{h}$. 
\end{itemize}
We emphasise that the physics of both options, (1) and (2), is captured by
\eqref{eq:gravNc} and is identical. Which part of the scaling of the
propagator is captured by $\mu$ and which one by $Z_{h}$ is
determined by the projection procedure. Note that the latter is also
approximation dependent.

In summary the coupled Yang-Mills--gravity system approaches the large $N_c$
limit via \eqref{eq:gravNc}. Whether or not this is seen in the current
approximation with the cutoff choice \eqref{eq:bound} is a technical
issue. If the approximation admits option (1) then the fixed point can
be approached, if (2) or a mixture of (1) and (2) is taken then the fixed
point cannot be seen due to the regulator bound in our setup. 
We emphasise again that this does not entail
the non-existence of the fixed point, which is guaranteed by the analysis of 
\autoref{sec:as-af}. The analysis here evaluates the capability of the
approximation to capture this fixed point. The understanding of this
structure and guaranteeing this capability of the approximation is of
chief importance when evaluating the stability of more complex
matter-gravity systems with genuine matter self-interaction: no
conclusion concerning the stability of these systems can be drawn if
the capability problem for the free-matter--gravity systems is not
resolved. Moreover, even if the fixed points exist, their physics may
be qualitatively biased by this problem.

\begin{figure*}[t!]
\includegraphics[width=\textwidth]{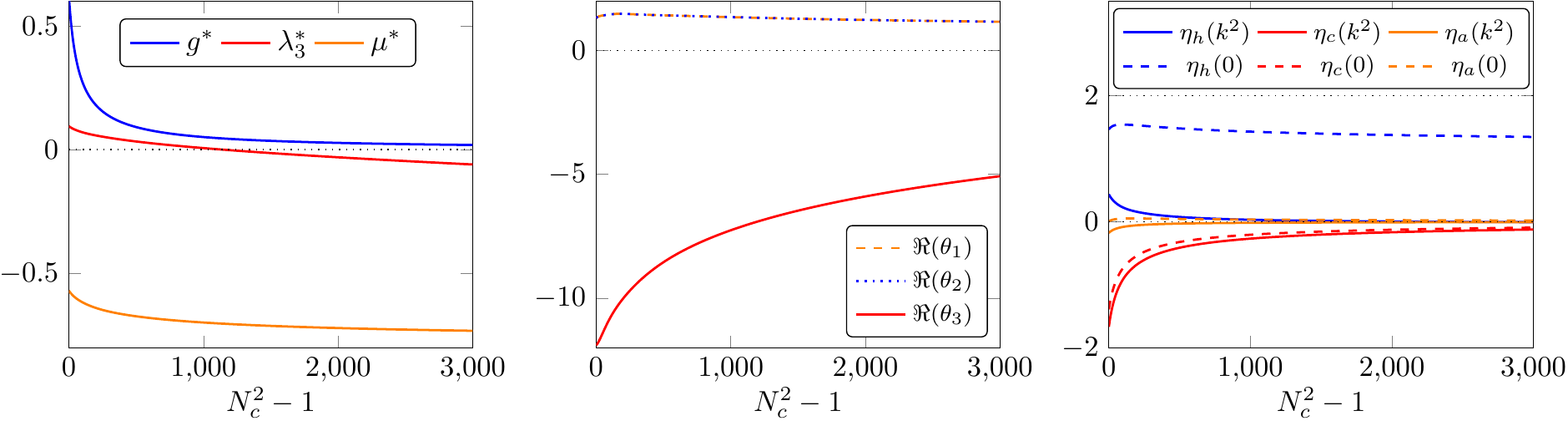}
\caption{Properties of the UV fixed point as a function of $N_c^2-1$
  in the uniform approximation with one Newton's coupling and
  with $c_{\mu,a}=\01{24\pi}\approx 0.0133$.  Displayed are the fixed
  point values (left panel), the critical exponents (central panel),
  and the anomalous dimensions (right panel).}
\label{fig:FPs-uniform-cmua}
\end{figure*}

\begin{figure*}[t!]
\includegraphics[width=\textwidth]{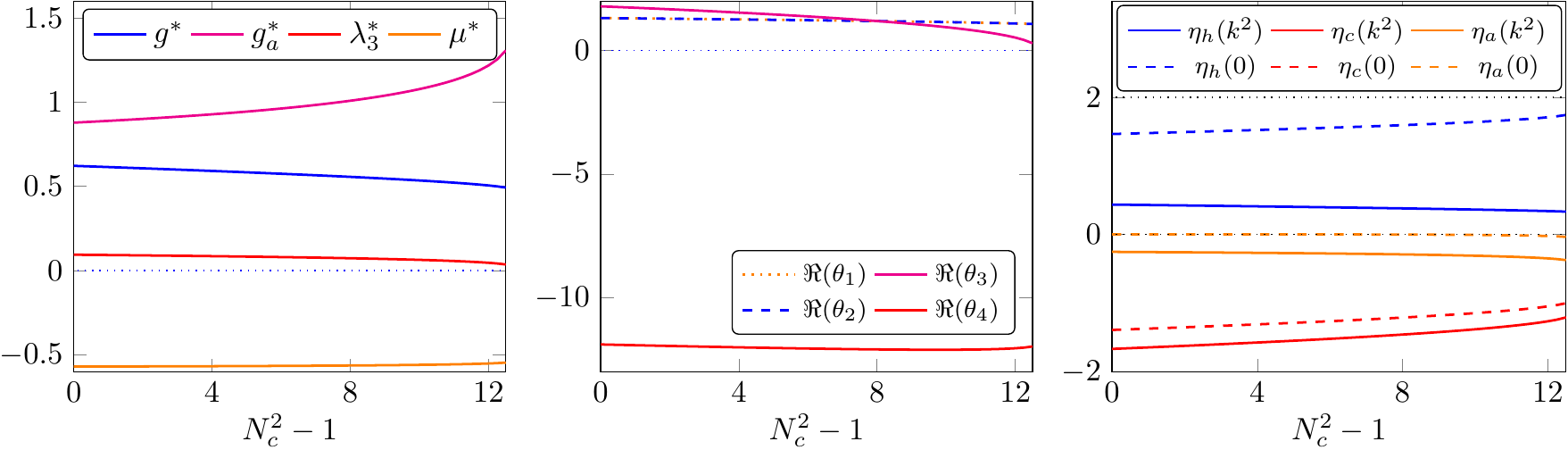}
\caption{Properties of the UV fixed point as a function of $N_c^2-1$
  in the approximation with two Newton's couplings and with the flat regulator, $c_{\mu,a}=0$.  
  Displayed are the fixed point values (left panel), the critical exponents
  (central panel), and the anomalous dimensions (right panel).}
\label{fig:FPs-2g}
\end{figure*}

\begin{figure*}[t!]
\includegraphics[width=\textwidth]{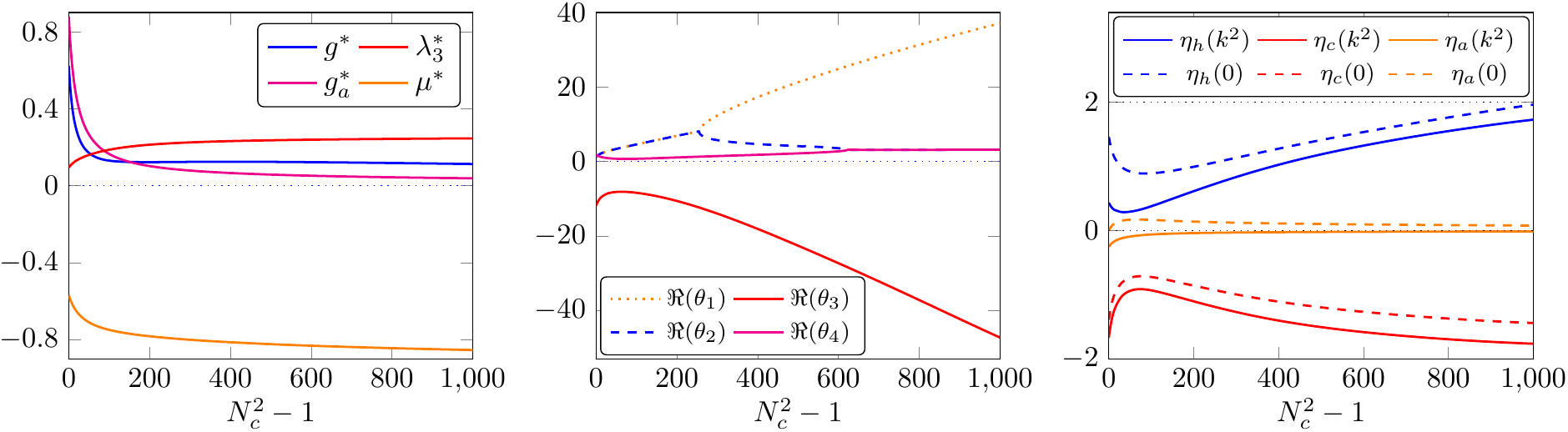}
\caption{Properties of the UV fixed point as a function of $N_c^2-1$
  in the approximation with two Newton's couplings and with 
  $c_{\mu,a}=\01{4\pi}\approx 0.08$.  Displayed are the fixed point
  values (left panel), the critical exponents (central panel), and the
  anomalous dimensions (right panel).}
\label{fig:FPs-2g-cmua}
\end{figure*}

\subsection{Results in the extended approximation}
\label{sec:extended-approx}
In the following analysis, we concentrate on the $g_a$ fixed point
equation and keep $g_c=g$. Before we extend the approximation to this
case, let us reevaluate the results with $g_a=g$ in the light of the
last \autoref{sec:dynscale}. There it has been deduced that a
consistent $N_c$ scaling requires $g^*\propto 1/N_c^2$ and either
\eqref{eq:1+mu} or \eqref{eq:etainf}, or both. \autoref{fig:FPs-uniform}
shows the consistent large $N_c$ scaling for Newton's coupling but
neither \eqref{eq:1+mu} nor \eqref{eq:etainf}. This comes as a surprise as
the system is asymptotically safe and the large $N_c$ limit in the
approximation $g=g_a$ is seemingly stable.  To investigate this
stability, we examine the regulator dependence of the coefficients of
the flow equations. To that end, we notice that the coefficients in the
$\mu$ equation (and the $g_3,g_a$ equations) are of crucial importance
for the stability of the system. The coefficient
$c_{\mu,a}=-1/(60\pi) \eta_a$ of the Yang-Mills contribution to the
graviton mass parameter is proportional to the gluon anomalous
dimension $\eta_a$: the leading coefficient vanishes, see
\eqref{eq:fullscale} and \eqref{eq:cmu}.  Indeed, choosing other regulators,
the leading order term is non-vanishing with
\begin{align}\label{eq:cmuarange} 
-0.2\lesssim c_{\mu,a}(R_k)\lesssim 0.2\,,
\end{align}
see App.~\ref{app:ym-sign-to-dot-mu}.
Typically, it supersedes the $\eta_a$-dependent term, and the flat
regulator appears to be a very special choice. If
$c_{\mu,a}\gtrsim 0.013$, we indeed find a solution, which is consistent
with \eqref{eq:1+mu}, see \autoref{fig:FPs-uniform-cmua} for
$c_{\mu,a}=\01{24\pi}\approx 0.0133$. 
In turn, for $c_{\mu,a}\lesssim -0.005$, we find solutions with
growing $\eta_h$, hence in the class \eqref{eq:etainf}. Accordingly, this
solution is not trustworthy with $\eta_h$ beyond the bound
\eqref{eq:bound}. Its failure simply is one of the approximation (within
this choice of regulator) rather than that of asymptotic safety.

In summary this leads us to a classification of the regulators
according to the large $N_c$ limit: they either induce the dynamical
readjustment of the scales via \eqref{eq:1+mu} or via \eqref{eq:etainf} or
they fall in between such as the flat cutoff. Within the current
approximation it is required that the readjustment happens via \eqref{eq:1+mu}. 

Now we are in the position to discuss the general case with
$g_a\neq g$. An optimal scenario would be that the inclusion of the
$g_a$ equation already stabilises the system such that it enforces the
dynamical readjustment via \eqref{eq:1+mu} for all regulators
proportional to $Z_h$. However, as we shall see, the general scheme
from the uniform approximation persists with this upgrade of
the approximation.

\subsubsection{No apparent \texorpdfstring{$N_c$}{Nc} scaling 
for \texorpdfstring{$\mu$}{mu} and \texorpdfstring{$\eta_h$}{eta h}}
\label{sec:complex}
In the uniform approximation with one Newton's coupling \eqref{eq:gn}, this scenario
was taken with regulators with $-0.005 \lesssim c_{\mu,a} \lesssim 0.013$.
A typical regulator in this class is the flat regulator used in the
present work. This scenario does not enhance the graviton propagator and hence,
does not fulfil \eqref{eq:gravNc}. The stability of the results 
in the large $N_c$ limit in the uniform approximation must thus rather
be considered a mere coincidence. Indeed in the extended truncation with $g\neq g_a$, 
the enhancement of the graviton propagator is not triggered by the included
$g_a$ equation, and consequently, the flat regulator does not have a stable 
large $N_c$ limit anymore. The fixed point values, critical exponents, and the 
anomalous dimensions in this approximation are shown in \autoref{fig:FPs-2g}.
The fixed point values show a
marginal $N_c$ dependence up to the point where the fixed point
vanishes into the complex plane at $N_c^2\approx13.5$, which is signalled by one of the
critical exponents going towards zero. The vanishing critical exponent can be
associated with $g_a$. Typically, this is interpreted as a
sign for the failure of asymptotic safety. Here it is evident that the
truncation cannot accommodate the dynamical readjustment of the scales
that takes place in the full system.  This could also signal an over-complete
system: $g$ and $g_a$ are related by diffeomorphism invariance. In any
case, the failure of the approximation can either lead to the
divergence of the couplings [related to \eqref{eq:etainf}], or in complex
parts of the fixed point values. For the flat regulator, the latter
scenario is taken.

\subsubsection{Scenario with \texorpdfstring{$1+\mu\propto 1/N_c^2$}{1 + mu ~ 1/(Nc*NC)}}
\label{sec:1+mu}
This scenario requires regulators with
$c_+< c_{\mu,a} <c_{\text{\tiny{max}}}$.  A typical regulator in this
class is the sharp regulator, see \eqref{eq:sharp} and \autoref{fig:regulator-scheme}.
Here, we do not
present a full analysis of this case but only change the coefficient
$c_{\mu,a}$ accordingly. This is justified in terms of linear small
perturbations of the system: $c_{\mu,a}$ is the only leading order coefficient in
the system that exhibits a qualitative change when changing the
regulator away from the flat regulator. Note however, that this
change ceases to be small for large $N_c$ as $c_{\mu,a}$ is
multiplied by $N_c^2$. If accompanied by a respective change of the
relative cutoff scales $k_h/k_a$, this factor could be
compensated. Then, however, we are directly in the stable regulator
choice with \eqref{eq:kh-ka}. Here, we are more interested in the
dynamical stabilisation and we refrain from the rescaling. The system
exhibits the $1/N_c^2$ scaling in the Newton's couplings,
$g^*$ and $g^*_a$, as well as the mass parameter $\mu^*$, see
\autoref{fig:FPs-2g-cmua} for $c_{\mu,a}\approx0.08$.
However, with this choice, the critical exponents of the fixed point
become rather large. We determined the constant $c_+ \approx 0.07$.

\subsubsection{Scenario with \texorpdfstring{$\eta_h$}{eta h} growing large }
\label{sec:etainf}
This scenario requires regulators with 
$-c_{\text{\tiny{min}}}< c_{\mu,a}<-c_-$. 
A typical regulator in this class is the exponential regulator, see
\eqref{eq:exp} and \autoref{fig:regulator-scheme}. For this class of regulators, both couplings grow
large, and we have the scenario with \eqref{eq:etainf} bound to fail to
provide fixed point solutions beyond a maximal $N_c$ due to the
failure of the approximation scheme.

\subsection{Resum\'{e}: Signatures of asymptotic safety of Yang-Mills--gravity systems}
\label{sec:as-discussion}
In summary, with the choice of the regulator, we can dial the different
scenarios that all entail the same physics: the dynamical
readjustment of the respective scales in the gauge and gravity
subsystems and the asymptotic safety of the combined system. The two
different scenarios are described in 
\autoref{sec:1+mu} and \autoref{sec:etainf}. Both scenarios entail the
same physics mechanism: the enhancement of the graviton propagator, see
\eqref{eq:gravNc}. This triggers the dominance of gravity
in the ultraviolet, which is clearly visible in the consecutive
integrating out of degrees of freedom discussed in
\autoref{sec:as-af}. The crucial property for the validity of this
structure is the asymptotic freedom of the Yang-Mills system, and
hence, the existence of the gauge system in a given background. This
property is trivially present in systems with free matter coupled to
gravity, and hence the present analysis extends to these cases.

This leaves us with the question of how to reevaluate the existing
results on matter-gravity system in the light of the present
findings. We first notice that the helpful peculiarity of the
Yang-Mills--gravity system that allowed us to easily access all the
different scenarios, is the possibility to choose the sign of
$c_{\mu,a}$ with the choice of the regulator. Clearly, the gauge
contribution to the running of the graviton mass parameter plays a
pivotal r\^ole for how the enhancement of the graviton
propagator in \eqref{eq:gravNc} is technically achieved. In the other
matter-gravity system this parameter has a definite sign, which is why
one sees a specific scenario for typical regulators. Collecting all
the results and restricting ourselves to truncations that resolve the
difference between fluctuation and background fields,
\cite{Meibohm:2015twa}, we find the following: 
\begin{itemize} 
\item[(1)] Fermion-gravity systems: they fall into the class
  \autoref{sec:1+mu}, and the asymptotic safety of the system can be
  accessed in the approximation. The required large flavour
  $N_f$ pattern with \eqref{eq:1+mu} is visible in the results.
\item[(2)] Scalar-gravity systems: they fall into the class \autoref{sec:etainf}, 
  and for large enough number of scalars $N_s$, the fixed point
  seemingly disappears due to the fixed point coupling $g^*$ and
  anomalous dimension $\eta_h$ growing too large. 
\item[(3)] Vector-gravity/Yang-Mills--gravity systems: this system
  has been discussed here, and it falls into all classes,
  \autoref{sec:complex}, \autoref{sec:1+mu} and
  \autoref{sec:etainf}. This also includes the $U(1)$ system.
\item[(4)] 
  Self-interacting gauge-matter--gravity systems: these
  systems only fall into the pattern described in
  \autoref{sec:complex}, \autoref{sec:1+mu}, and \autoref{sec:etainf}
  if the gauge-matter system is itself ultraviolet stable. For
  example, one flavour QED exhibits a UV-Landau pole and is stabilised
  by gravity, which makes the combined system asymptotically safe, for
  a comprehensive analysis see
  \cite{Christiansen:2017gtg,Christiansen:2017qca}. Adding more
  flavours potentially destabilises the system; however, such an
  analysis has to avoid the interpretation of the seeming failure of
  asymptotic safety described here. One possibility to take this into
  account is the scale adjustment \eqref{eq:kh-ka}. This discussion also
  carries over to general gauge-matter--gravity systems including the
  Standard Model and its extensions.
\end{itemize} 

In summary, this explains the results obtained in gravitationally
interacting gauge-matter--gravity systems, which are the basis of
general gauge-matter--gravity system. While it suggests the use of
relative cutoff scales such as \eqref{eq:kh-ka}, it still leaves us with
the task of devising approximations that are capable of capturing the
dynamical readjustment of scales that happens in gravitationally
interacting gauge-matter--gravity systems. In particular, the
marginal operator $R^2 \ln( 1+ R/{k_a^{\text{\tiny IR}}}^2)$,
cf.~\eqref{eq:R2logR}, has to be included as discussed in
\autoref{sec:ASgluegrav}.

Besides this task, the present analysis also requires a careful
reanalysis of phenomenological bounds on ultraviolet fixed point
couplings. It is well-known that the values of the latter are subject
to rescalings and only dimensionless products of couplings such as
$g^*\lambda^*$ possibly have a direct physical interpretation. We have
argued here that the dynamically adjusted or explicitly adjusted
relative cutoff scales ask for a reassessment also of these
dimensionless products.

\section{Summary and conclusions}
\label{sec:summary}
We have investigated the prospect for asymptotic safety of gravity in
the presence of general matter fields. A main new addition are general
arguments, which state that if matter remains sufficiently weakly
coupled in the UV, or even free, asymptotic safety for the combined
matter-gravity theory follows, in essence, from asymptotic safety of
pure gravity (\autoref{sec:as-af}).  Ultimately, the UV dominance of
gravitons relates to the fact that the integrating out of UV-free
matter fields only generates local counter terms in the gravitational
sector. 

Our reasoning has been tested comprehensively for Yang-Mills theory
coupled to gravity. Using identical cutoffs for gravity and matter,
we invariably find that asymptotic safety arises
at a partially interacting fixed point with asymptotic freedom in the
Yang-Mills and asymptotic safety in the gravity sector. Fluctuations
of the gravitons dominate over those by matter fields including in the
asymptotic limit of infinite $N_c$
(\autoref{fig:FPs-uniform}). 
Interestingly, the UV dominance of
gravity can materialise itself in different manners
(\autoref{fig:FPs-uniform-cmua},~\ref{fig:FPs-2g},~\ref{fig:FPs-2g-cmua}),
strongly depending on technical parameters of the theory such as the
gauge, the regularisation, and the momentum cutoff. The overall
physics, however, is not affected
(\autoref{fig:regulator-scheme}). This pattern is reminiscent of how
confinement arises in gauge-fixed continuum formulations of QCD.
It is also worth noting that the observed $N_c$ independence
with identical cutoffs follows automatically, if, instead,  
 "relative cutoffs"  for  matter and gravity fluctuations are adopted,  following \eqref{eq:kh-ka}. 
This may prove useful for  practical studies of gravity-matter systems in set approximations. 
The necessity for "relative  cutoffs" 
is well-understood in  condensed matter systems, 
albeit for  other reasons \cite{Diehl:2007xz,Pawlowski:2015mlf}.

There are several points that would benefit from further study in the
future. While we explained in general terms how findings extend to
more general matter sectors (\autoref{sec:general}), it would seem
useful to further substantiate this in explicit studies. Also, our study 
highlighted the appearance of logarithmic terms such as $R^2\ln R$, 
and similar (\autoref{sec:as-af}). These classically marginal terms are
of relevance for the question of unitarity of
asymptotically safe gravity. It remains  to be seen 
whether they affect the  observed $N_c$ independence of gravity-matter 
fixed points in any significant manner (\autoref{sec:general}).
Finally, our findings offer a natural reinterpretation of earlier results. 
It is important to confirm whether this is sufficient to remove a tension 
amongst previous findings based on different implementations of the 
renormalisation group. Understanding these aspects opens a door
towards reliable conclusions for UV completions of the Standard Model
or its extensions.

\vspace{.5cm}
\centerline{\bf Acknowledgements} 
The authors thank A.~Bonanno,
A.~Eichhorn, H.~Gies S.~Lippoldt, and C.~Wetterich for discussions.
MR acknowledges funding from IMPRS-PTFS.  This work is supported by
EMMI, the grant ERC-AdG-290623, the DFG through grant EI 1037-1, the
BMBF grant 05P12VHCTG, and is part of and supported by the DFG
Collaborative Research Centre "SFB 1225 (ISOQUANT)".

\appendix

\section{Regulators}
\label{app:regulators}
In the present work we use the optimised or flat regulator 
\cite{Litim:2000ci,Litim:2001up,Litim:2001fd,Litim:2006ag} for all field
modes. Specifically, the superfield regulator at $\bar g=\id$ and
$\bar A=0$ with flat Euclidean background metric is given by
\begin{align}\nonumber  
 R^{ij}_k(p) &= \delta^{ij}\,
 \left. \Gamma^{(\phi_i \phi^*_i )}(p)\right|_{\mu=0} r_{\phi_i}(p^2/k^2)\,,\\
\qquad 
r(x) &=\left(\01x -1\right) \theta(1-x)\,.
\label{eq:flat}
\end{align} 
Here, $\phi^*$ is the dual superfield with
$\phi^*= \left(h_{\mu\nu},-\bar{c}_\mu,c_\mu,A_\mu,-\bar{c},c\right)$.
The regulator \eqref{eq:flat} is diagonal in field space keeping in
mind the symplectic metric and allows for analytic expressions of the flow \cite{Litim:2003vp}.
For the general scaling analysis we also discuss more general regulators, in particular, 
we refer to the exponential regulator with 
\begin{align}\label{eq:exp}
r(x) &= \01{\exp(x)-1}\,,
\end{align}
and to the sharp cutoff regulator with 
\begin{align}\label{eq:sharp}
r(x) &= \01{\theta(x-1)}-1\,.
\end{align}
These regulators and variants thereof can be used to scan the space of cutoff functions \cite{Litim:2002cf,Litim:2007jb}.

\section{Regulator dependence of the gluon contribution to the graviton mass parameter}
\label{app:ym-sign-to-dot-mu}
The coefficient $c_{\mu,a}$, which parameterises the gluon contribution to the graviton mass parameter, is given by
\begin{align}
 c_{\mu,a} &=-\0{\text{Flow}_a^{(2h)}(p^2=0)}{g (N_c^2-1)} \notag\\
 &= \01{3\pi} \int\!  \0{\mathrm d x \,x\, \dot r_h(x)}{(1+r_h(x))^2} \left( \04{1+r_h(x)} -3  \right) , \label{eq:kinematic-id}
\end{align}
with $x=\0{q^2}{k^2}$, $\eta_a=0$ on the right-hand side and where the angular integration was already performed.
We now use that
\begin{align}
 k\partial_k r_h(k,x) = k \0{\partial x}{\partial k} \partial_x r_h(k,x) = -2 x \partial_x r_h(k,x) \,,
\end{align}
and consequently we get
\begin{align}
 c_{\mu,a} ={}&  -\0{2}{3\pi} \int\! \mathrm d x \,x^2 \left( \partial_x\left(\02{(1+r_h(x))^2} -2 \right)\right. \notag\\ 
 &\hspace{.5cm}\left.-\partial_x\left(\03{1+r_h(x)} -3 \right) \right)\,,
\end{align}
where we added zeros in order to perform the partial integration without boundary terms.
The result after partial integration is
\begin{align}
 c_{\mu,a} ={}& \0{4}{3\pi} \int \!\mathrm d x \,x  \0{r_h(x)(r_h(x)-1)}{(1+r_h(x))^2}\,.
\end{align}
We have evaluated this integral for different types of regulator shape functions.
The results are displayed in \autoref{tab:gluon-to-dot-mu}.
The flat regulator evaluates this integral to zero, while exponential regulators give a positive sign 
and step-like or sharp regulators even give a negative sign.
The usual expectation is that the regulator changes the size of a contribution but not its sign.
In this case, however, two diagrams cancel each other approximately 
and by changing the regulator, we shift the weights between these two diagrams.
Thus, any sign of this contribution is possible.

\begin{table}[t!]
\caption{Gluon contribution to the graviton mass parameter for different regulators.
Remarkably, the contribution does not only change in size but also its sign.}
\label{tab:gluon-to-dot-mu}
\begin{tabular}{c|c}
Regulator & $c_{\mu,a}$ \\[.1cm]
\hline\\[-.2cm]
$r(x)= \01{\exp(x)-1}$ & $-0.21$  \\[.1cm]
\hline\\[-.2cm]
$r(x)= \01x \exp(-x^2)$ & $-0.027$  \\[.1cm]
\hline\\[-.2cm]
$r(x)=(\01x-1)\Theta(1-x)$ & $0$  \\[.1cm]
\hline\\[-.2cm]
$r(x)=\01x\Theta(1-x)$ & $0.034$  \\[.1cm]
\hline\\[-.2cm]
$r(x)=\0{10}x\Theta(1-x)$ & $0.17$  \\[.1cm]
\hline\\[-.2cm]
$r(x)=\01{\Theta(x-1)}-1$ & $\0{2}{3 \pi}\approx0.21$  
\end{tabular}
\end{table}

\begin{figure*}[t!]
\includegraphics[width=.9\hsize]{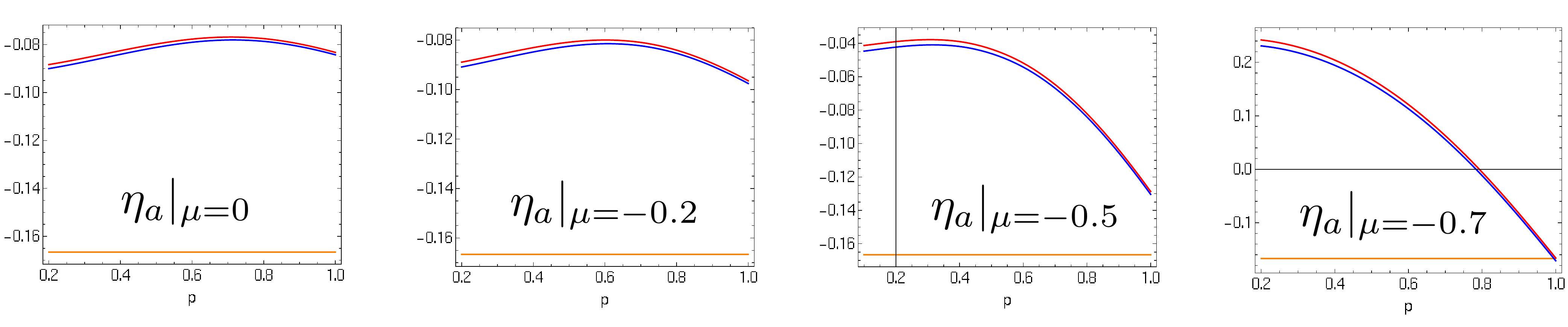}
\vskip-.3cm
\caption{Shown is the momentum dependence of the graviton contribution to the
gluon anomalous dimension $\eta_a$ for different values of the graviton mass
parameter $\mu=0,\,-0.2,\, -0.5$, and $-0.7$ (from left to right).
In each case, starting with a flat trial function (orange), a fast convergence from  
first (blue) to second (red) order in the iteration \eqref{eq:Fredholm_equ_general_order} 
is observed ($g=0.5$ and $\eta_h=0.5$).}
   \label{fig:etaA_p}
\end{figure*}

\section{Inhomogeneous Fredholm integral equations of the second kind}
\label{app:integral_equ}
In this Appendix, we discuss methods to solve Fredholm integral equations
on the example of the gluon anomalous dimension
\begin{align}
\eta_a(p^2) = f(p^2) + g \int\! \0{\mathrm d^4q}{(2 \pi)^4}
\,K\!\left(p,q,\mu, \eta_h \right) \eta_a(q^2) \,,
\label{eq:Fredholm_equ_app}
\end{align}
see \autoref{sec:gravity-to-YM}.
Fredholm integral equations of the second kind are a well-known topic in pure
and applied mathematics and there are several methods in order to
solve such equations. 
A straightforward numerical solution is the so-called Nystroem method that is
based on discretisation of the integral operator with quadratures on $N$ points.
By doing so, one obtains Riemann sums that reduce to a system of $N$ linear
equations. Moreover,  if there exist a solution to \eqref{eq:Fredholm_equ_app}, it can 
be shown by the general theory of such equations
that it is unique and the discretised version converges towards this solution
in the limit $N \to \infty$. 
Another method that comes along with less numerical effort are iterative
solutions based on the resolvent formalism and the Liouville-Neumann series.
The basic idea of this approach is as follows. In order to get a feeling for
such integral equations, we observe that for $g=0$, the
unique solution to \eqref{eq:Fredholm_equ_app} is trivially given by the
inhomogeneity $f(p^2)$. Hence, if $g$ is small in some sense, it seems
reasonable that $f(p^2)$ is at least a good zeroth order approximation to
the full solution $\eta_a(p^2)$, i.e.\ $\eta_a(p^2) \approx \eta_{a,0}(p^2) 
\equiv
f(p^2)$.
In a first iteration step, we substitute $\eta_{a,0}(q^2)$ for $\eta_a(q^2)$ 
under
the integral on the right-hand side of the integral equation
\eqref{eq:Fredholm_equ_app},
\begin{align}
\eta_{a,1}(p^2) = f(p^2) + g \int \0{\mathrm d^4q}{(2 \pi)^4}
K\left(p,q,\mu,\eta_h\right) \eta_{a,0}(q^2) \,.
\label{Fredholm_equ_1st_order} 
\end{align}
In this spirit we can construct iteratively a sequence
$\left(\eta_{a,i}(p^2)\right)_{i \in \mathbb{N}}$ with
\begin{align}
\eta_{a,i+1}(p^2) = f(p^2) + g \int \0{\mathrm d^4q}{(2 \pi)^4}
K\left(p,q,\mu,\eta_h\right) \eta_{a,i}(q^2) \,.
\label{eq:Fredholm_equ_general_order} 
\end{align}
The convergence properties depend on the kernel $K$ and the coupling constant
$g$. We observe that due to the regulator structure, the kernel $K$ is
proportional to $r_a(q^2)$. Therefore, the kernel is
integrable with respect to the loop momentum $q$. For the sake of simplicity,
we will assume in the following  a flat regulator $r_a(q^2) \sim
\theta(1-q^2)$, where $q$ is the dimensionless momentum. The
discussion can be generalised straightforwardly to arbitrary regulators. With
a flat regulator, we write $K(p,q)=: \theta(1-q^2) \check{K}(p,q)$.
As a consequence, the integral in the Fredholm equation is defined on the domain
$[0,1]$, and in all equations, $K$ is substituted by $\check{K}$.
Moreover, we define the angular averaged kernel 
\begin{align}
\langle  \check{K} \rangle_\Omega (p,q,\mu,\eta_h) := \int_{S^3} 
\0{\mathrm{d} \Omega}{(2 \pi)^4} \check{K}(p,q,x,\mu,\eta_h)\, ,  
\end{align}
where $\mathrm{d} \Omega$ is the canonical measure on the three sphere.
The kernel
$\langle  \check{K} \rangle_\Omega$ can be normed,
in particular, it exists its $2$-norm with respect to the first two arguments
\begin{align}
\left| \left| \langle  \check{K} \rangle_\Omega \right| \right|_2 := \left( 
\int_0^1 \!\! \int_0^1 \!\!
\mathrm{d}q \, \mathrm{d}p
\left| \langle  \check{K} \rangle_\Omega (p,q,\mu,\eta_h) \right|^2 
\right)^{1/2}
\end{align}
It can then be shown that the sequence $\left( \eta_i(p^2) \right)_{i \in 
\mathbb{N}}$ 
converges towards the full solution, i.e., 
\begin{align}
\lim_{i \rightarrow \infty} \eta_{a,i}(p^2) = \eta_a(p^2) \,, 
\end{align}
if the kernel is bounded as 
\begin{align}
\big| g \big| \, \, \left| \left| \langle  \check{K} \rangle_\Omega \right| 
\right|_2 < 1 \, .
\label{eq:kernel_bound}
\end{align}
The solution can then be written as a Liouville-Neumann series according to
\begin{align}
\eta_a(p^2) =  f(p^2) +  g \int_{\mathbb{R}^4}  \0{\mathrm d^4q}{(2 \pi)^4}
R\left(p,q,\mu,\eta_h,g \right) f(q^2) \, ,
\end{align}
with the resolvent kernel 
\begin{align}
R\left(p,q,\mu,\eta_h,g \right) = \sum_{i=1}^{\infty} g^{i-1} 
K_i\left(p,q,\mu,\eta_h \right) \, ,
\end{align}
where $K_i$ are the iterated kernels given by
\begin{align}
\notag & K_i\left(p,q,\mu,\eta_h \right) = \int
\int \dots
\int\0{\mathrm d^4q_1}{(2 \pi)^4} \0{\mathrm d^4q_2}{(2 
\pi)^4}
\dots \0{\mathrm d^4q_{i-1}}{(2 \pi)^4}  
\\ \notag & \times K\left(p,q_1,\mu,\eta_h \right)
K\left(q_1,q_2,\mu,\eta_h \right) \times \dots\,  
\\ & \times \, K\left(q_{i-1},q,\mu,\eta_h \right) \,.
\end{align}
By truncating the resolvent series at some finite order $i_0$, one
obtains an approximate solution to the integral equation. If the
bound \eqref{eq:kernel_bound} is satisfied, the Liouville-Neumann series
converges for any smooth initial choice $\eta_{a,0}$. One can also
choose zeroth iterations that are different from the inhomogeneity
$f(p^2)$. It is clear that convergence properties depend on the
initial choice. For instance, if one has the correct guess for the full
solution and uses this as a starting point for the iteration, then
one finds $\eta_{a,0} = \eta_{a,1}$, and one can conclude that the
exact solution has been found.  Additionally, there are improved
iteration schemes that increase the radius of convergence
significantly. In \cite{buckner1948special}, it has been proven that it
exists a parameter $c \in \mathbb{R}$, such that the iteration
prescription
\begin{align}
\eta_{a,i+1}(p^2) ={}&  (1-c)f(p^2) + c \, \eta_{a,i}(p^2) \\ 
& + (1-c) \, g \int \0{\mathrm d^4 q}{(2 \pi)^4}
K\left(p,q,\mu,\eta_h \right) \eta_{a,i}(q^2) \, \notag
\end{align}
has a radius of convergence that is larger than the one of the
standard Liouville-Neumann series, which is obtained from the improved
iterations with $c=0$.

The convergence in the present system is analysed in \autoref{fig:etaA_p}.
We plot $\eta_a(p^2)$ for some specific
parameter values. All these plots are obtained for $g=0.5$; however, we stress
that the sign of $\eta_a$ does not depend on this choice as the result is a power series in
$g$. We investigate the iterations, where we have always assumed a
constant function $\eta_{a,0} = \mathrm{const}$ as a first
approximation. We then plot the first, second, and third order and find rapid convergence in all
cases, which is expected as we have checked that the kernel in \eqref{eq:RHS_YM} generates
a very large radius of convergence. The third iteration is for this choice of
$\eta_{a,0}$ not even visible any more, since the
corresponding curve lies exactly on top of the second iteration.

\section{Sign of the gluon anomalous dimension}
\label{app:sign-change}
In this Appendix, we discuss the stability of the sign of the gluon anomalous dimension.
As discussed in \autoref{sec:QG_to_YM}, we need a negative sign in order to obtain asymptotic freedom in the gauge sector.
This directly corresponds to the demand that the gravity contributions to the gluon anomalous dimension should be negative.
In the App.~\ref{app:integral_equ} we discussed the full momentum dependent solution of $\eta_a(p^2)$.
We further argued in \autoref{sec:QG_to_YM} that the sign at $p^2=k^2$ is the decisive one for the Yang-Mills beta function.
In the following sections, we present different approximations to the gluon anomalous dimension, and how stable the sign is within these approximations.

\begin{figure*}[t!]
\includegraphics[width=.9\hsize]{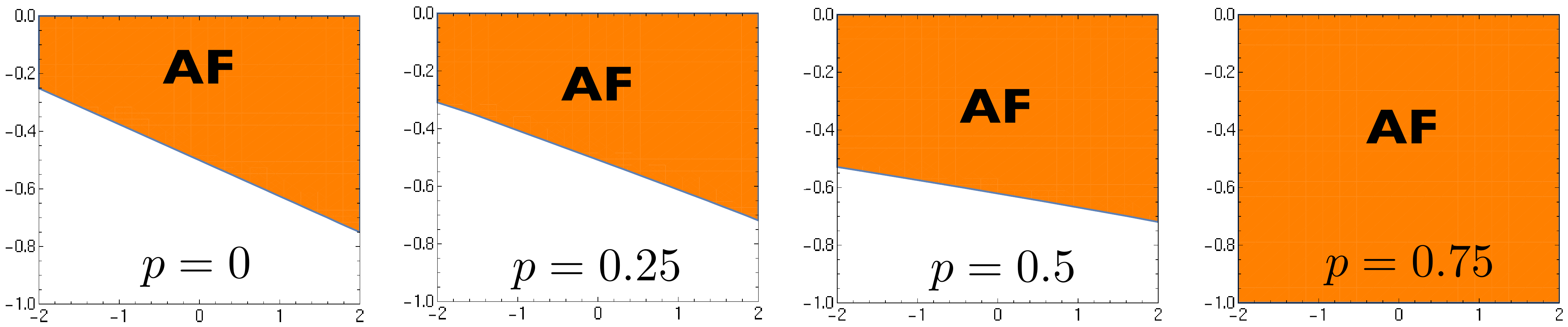}
\caption{In the plane of the graviton anomalous dimension ($\eta_h^*$, lower axis) and the graviton mass parameter $\mu^*$,
the region  with asymptotic freedom (AF) is coloured (orange) corresponding to a positive sign of the gluon anomalous dimension $\eta_a$. 
Moreover, the gluon anomalous dimension is determined from a momentum derivative evaluated  at different momenta $p=0,\, 0.25,\, 0.5$, and $0.75$ (from left to right).
The domain with asymptotic freedom consistently grows as soon as momenta of order of the RG scale are adopted.}
\label{fig:sign_change}
\end{figure*}

\subsection{Derivative at vanishing momentum}
\label{app:der_zero}
The simplest approximation is to assume a momentum independent anomalous dimension and
to obtain an equation for $\eta_a$ with a derivative at $p^2=0$.
The equation for $\eta_a$ is then given by
\begin{align}
\eta_{a,h}= - \partial_{p^2} \text{Flow}^{(AA)}_h \bigg|_{p^2=0}\,.
\end{align}
We obtain the analytic result
\begin{align}
\eta_{a,h}= -  \0{g}{8 \pi} \left(
\0{8-\eta_a}{1+\mu} -  
\0{4-\eta_h}{(1+\mu)^2} \right) \,,
\end{align}
which is identical to the $\eta_a$ in the UV if the gauge sector is asymptotically free.
Therefore, assuming a fixed point in the gravitational sector, we are left with the ultraviolet limit
\begin{align}\label{andim_UV}
\eta_a^* = 
\0{g^*}{1-\0{g^*}{8 \pi (1+ \mu^*)}} \left( \0{4+8 \mu^*+\eta_h^*}{8 \pi (1+\mu^*)^2} \right) \, .
\end{align}
This function changes sign at the critical value
\begin{align}
\label{eq:sign_change}
\mu^*_{\mathrm{crit}} = -\0{1}{8}(4+\eta_h^*) \, .
\end{align}
Moreover, there is a pole at $\mu^* = -1 + \0{g^*}{8 \pi}$ with
another sign change for the regimes to the left and to the right of
the pole. However, this sign change at the pole can be neglected, as
usual fixed point values of $g$ are $\mathcal{O}(1)$. For fixed point
values of this order, the pole is located at $\mu^* \approx -0.96$,
which in turn is a fixed point value that is very unusual.  Therefore,
we assume the overall prefactor in \eqref{andim_UV} to be positive.
Then, $\eta_a^* \gtrless 0$ for
$\mu^* \lessgtr -\0{1}{8}(4+\eta_h^*)$.  This agrees with previous
computations in the background field approximation, where
$\eta^*_{h} =-2$ and $\mu = -2 \lambda$, and consequently,
$\lambda^*_{\mathrm{crit}} = \0{1}{8}$ \cite{Folkerts:2011jz}.  In our
more general case, the anomalous dimension of the graviton is not
fixed by the fixed point condition for Newton's coupling. The fixed
point value for the graviton mass parameter where the gravitational
contribution changes sign is plotted against the graviton anomalous
dimension in the left panel of \autoref{fig:sign_change}. There are some bounds on
anomalous dimensions for well-defined theories. From previous results
\cite{Christiansen:2012rx,Christiansen:2014raa,Christiansen:2015rva,%
Meibohm:2015twa,Christiansen:2016sjn,Denz:2016qks},
we know that typical fixed point values are roughly given by
$\eta_h \approx 1$ and $\mu\approx -0.6$, which is just at the
critical value where asymptotic freedom is lost.

We conclude that in this simplest approximation the stability of asymptotic
freedom is not guaranteed, but depends strongly on subtle effects in the
gravity sector. In the following, we investigate how this picture changes in
more elaborate approximations and specifications.

\subsection{Derivative at non-vanishing momentum}
\label{app:der_gen}
We now generalise the procedure from the previous section and use a 
derivative at finite momentum. The equation for $\eta_a$ is then given by
\begin{align}
\eta_{a,h}= - \partial_{p^2} \text{Flow}^{(AA)}_h \bigg|_{p=\alpha k}\,.
\end{align}
For such derivatives the results are only numerical.
In \autoref{fig:sign_change} we show the results for $\alpha=\014,\012,\034$.
We again display the sign of the gluon anomalous dimension in the $(\mu^*,\eta_h^*)$ plane.
We find the encouraging result that the area, which does not support asymptotic freedom in the gauge sector 
is getting smaller with an increasing $\alpha$.
With a derivative at $p^2=k^2$, the region has completely disappeared from the investigated area.
We conclude that with this generalised derivation of the gluon anomalous dimension, 
asymptotic freedom is supported in the whole important parameter region of gravity.

\subsection{Finite differences}
\label{app:fd}
\begin{figure*}
\includegraphics[width=.9\hsize]{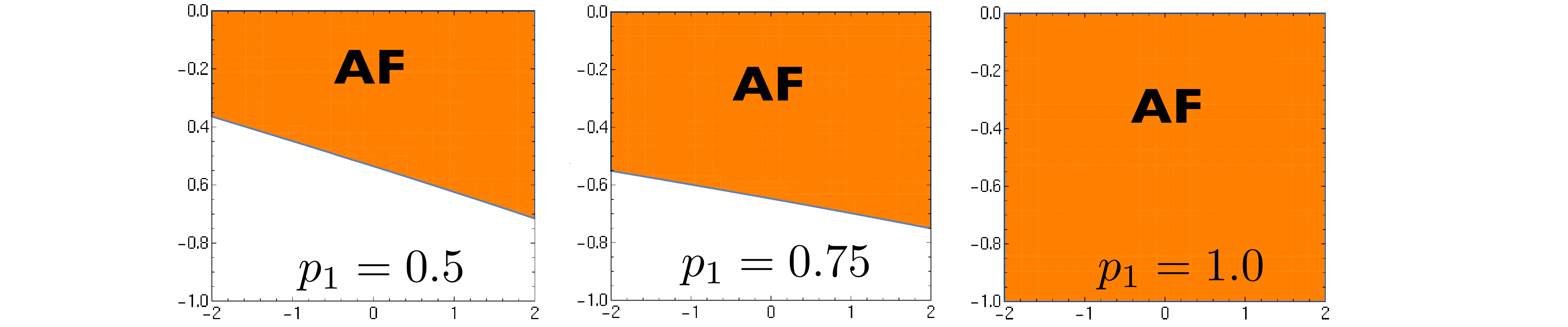}
\caption{
Same as \autoref{fig:sign_change}, except that 
the gluon anomalous dimension is determined from a finite difference 
derivative \eqref{eq:finite-diff} with $p_2=0$ and various momenta $p_1=\012, \034,1$ (from left to right).
The domain with asymptotic freedom consistently grows with growing
$p_1-p_2$ of the order of the RG scale, fully consistent with \autoref{fig:etaA_p}.}
\label{fig:sign_change_fd}
\end{figure*}

A further generalisation of the procedure from the previous sections
is to derive the gluon anomalous dimension by a finite difference.
In this case, we define $\eta_a$ to be momentum dependent.
It is then given by
\begin{align}
\label{eq:finite-diff}
\eta_{a,h} (p^2)= - \0{\text{Flow}^{(AA)}_h(p^2)-\text{Flow}^{(AA)}_h(0)}{p^2}\,.
\end{align}
The corresponding results are presented in \autoref{fig:sign_change_fd} for $p=\012,\034,1$.
The results are very similar to the ones with the derivative definition at non-vanishing momentum.
The gluon anomalous dimension is negative and supports asymptotic freedom
if we evaluate it at $p^2=k^2$.
This is also the approximation for $\eta_a$ that we utilise throughout this work
and also \autoref{fig:sign-change-main} is computed with this approximation.

\section{Scaling equations}
\label{app:scaling}
In this Appendix, we augment the analysis from \autoref{sec:general} 
by providing scaling equations for all couplings. 
In particular, we are lifting the identification \eqref{eq:gn}.
Here we extract the fixed point scaling from a flat
regulator choice and utilise a reparameterisation of the flow
equations that minimises the occurrence of factors of $1+\mu$.
Moreover, in the previous chapter, we have utilised
projections on gravitational couplings $g_n$ and $g_{aah}$ within a
finite difference construction. In the literature,
projections with derivatives at vanishing momentum, $p^2=0$, are
often used. It has been argued in
\cite{Christiansen:2012rx,Christiansen:2014raa,Christiansen:2015rva,
  Meibohm:2015twa,Christiansen:2016sjn,Denz:2016qks,Christiansen:2017gtg}
that this definition has large ambiguities at $p^2=0$, which limits its
applicability. Still, it has the charm of providing analytic
flows and fixed point equations and hence facilitating the access
to the current analysis.

The structure of the flow and fixed point equations is more
apparent if we absorb $1/(1+\mu)$-factors in the gravitational couplings with
\begin{subequations} \label{eq:bargn}
\begin{align}\label{eq:bargng}
  \bar g_n &= g_n\left( \0{1}{1+\mu}\right)^{\gamma_n}\,, 
  &
  \bar g_{\bar c c  h^n} &= g_{\bar c c  h^n} \left(\0{1}{1+\mu} \right)^{\gamma_{c}}\,, \notag\\ 
  \bar g_{a^n h^m} &= g_{a^n h^m} \left( \0{1}{1+\mu}\right)^{\gamma_a} \,,
\end{align}
with the scaling coefficients 
\begin{align}\label{eq:gamma} 
\gamma_n= \0{n}{n-2}\,,\qquad \qquad  \gamma_a=\gamma_c=1\,,
\end{align}
\end{subequations} 
and $\mu,\lambda_n$ are not rescaled. This removes all potentially
singular factors $1/(1+\mu)$-factors in the diagrams that stem from
the respective powers of the graviton propagators in the loops.
It still leaves us with contributions proportional to $1/(1+\mu)$ due to
the projection procedure with derivatives at $p^2=0$ and due to regulator insertions.
The rescaling power of $1/(1+\mu)$ varies between $1/(1+\mu)^3$ for the lowest coupling $g_3$ and $1/(1+\mu)$
for $g_{n\to\infty}$. 

In the following equations we identify blocks of
gravitational couplings: as before all gravitational self-couplings
$\bar g_n,\bar g_{\bar c c h^n}$ are identified with $\bar g_3$ and all
$\lambda_n$ are identified with $\lambda_3$. Additionally, we identify
all Yang-Mills--gravity interactions $\bar g_{aa h^n}$ with $\bar g_{aa h}$.
This leads us to
\begin{subequations}\label{eq:bargid}
\begin{align}
  \bar g_n= \bar g_3=\bar g\,,\qquad 
  \lambda_{n>2}=\lambda_3\,, 
  \label{eq:barghid}
\end{align}
for the pure gravity couplings and
\begin{align}
  \bar g_{\bar c c h^n}=\bar g_{c}\,,
  \qqquad 
  \bar g_{aa h^n}=\bar g_a \,,  
  \label{eq:bargaid}
\end{align}
\end{subequations} 
for the ghost-graviton and gluon-graviton couplings.
We emphasise that \eqref{eq:bargid} and \eqref{eq:bargng} imply
\begin{align}\label{eq:gnnecg3}
g_n=  g_3 (1+\mu)^{ \gamma_3-\gamma_n}\,, 
\end{align}
with $\gamma_3>\gamma_n$. Eq.\eqref{eq:gnnecg3} seemingly entails the
irrelevance of the lower order couplings $g_n$ for $\mu\to
-1$. However, the lower order couplings contribute to diagrams with
more graviton propagators. In combination, this leads to a uniform
scaling of all diagrams as expected in a scaling limit. Note that the
scaling analysis can also be performed if removing the approximation
\eqref{eq:bargid}. It leads to an identical scaling
$\bar g_n\sim \bar g_3$ and $\bar g_{aa h^n}\sim \bar g_{aa h} $. The
discussion of such a full analysis is deferred to future work.

Here, we are only interested in the relative scaling between the pure
gravity and Yang-Mills gravity diagrams, and simply discuss the
structure of these equations. To that end, we use the analytic pure
gravity equations derived in \cite{Christiansen:2015rva,Denz:2016qks} 
expressed with the
rescaled couplings \eqref{eq:bargn}. We also use the identification
\eqref{eq:bargid}, and additionally, we suppress the ghost contribution 
for simplicity. 
The ghost contribution comes with the same power in $1+\mu$ as the gluon contribution.
The analysis is facilitated by only using positive
coefficients $c_i,d_i$, making the relative
signs of the different terms apparent.
In general the sign of some of these coefficients depends on $\lambda_3$, 
and we define them such that they are positive at $\lambda_3=0$.
The explicit values for the
coefficients is provided in App.~\ref{app:coeffs}. Within this
notation, all factors $1/(1+\mu)$ in the loops are absorbed in the
couplings except the one, which comes from external momentum
derivatives of propagators, $\partial_{p^2} G$, due to the projection
procedure or from regulator insertions.
In summary, we are led to
\begin{subequations}
\label{eq:fullscale}
\begin{align}\nonumber 
  \dot  \mu={}& -( 2-  \eta_h) \mu 
            -  \bar g \left[  c_{\mu, h} + (1+\mu)(N_c^2-1)
              c_{\mu, a}\,\0{ \bar g_{a}^{\ }}{\bar g }
              \right]\,,\\  \nonumber 
  \dot {\bar g} ={}& \left( 2 + 
                   3\bar \eta_h  \right) \bar g\\\nonumber 
            &\,-\bar g^{2} \left[\0{c_{\bar g, h}}{1+\mu}+\0{d_{\bar g, h}}{(1+\mu)^2} + (N_c^2-1) c_{\bar g, a}
              \left( \0{ \bar g_{a} }{\bar g }
              \right)^{\032} \right]\,,\\\nonumber 
  \dot  \lambda_3={}&- \left( 1 +\0{\partial_t { \bar g}}{2 \bar g}-\032 \bar \eta_h  \right)\lambda_3\\ 
            &\,+\bar g\left[ \0{c_{\lambda_3, h}}{1+\mu} + 
              (N_c^2-1) c_{\lambda_3, a}
              \left( \0{ \bar g_{a} }{\bar g }
              \right)^{\032} \right]\,,\label{eq:fullscaleg}
\end{align}
for the pure gravity couplings. Here, the term $d_{\bar g,h}/(1+\mu)^2$
stems from the $\partial_{p^2} G$ contributions, and all coefficients
$c,d$ from graviton loops depend on $\lambda_3$ with $c(0), d(0)
>0$. The ghost-graviton and the gauge-graviton coupling have the flows
\begin{align} 
  \dot { \bar g}_{a}
  = &\,\left( 2+2 \eta_a +\bar\eta_h \right) \bar g_{a}^{\ }-
     \bar g_{a}^{2}\,\Biggl[ -c_{\bar g_a,a}+\0{d_{\bar g_a,a}}{1+\mu}  \notag\\
   &\,\hspace{.5cm}+\left( c_{\bar g_a,h}- \0{d_{\bar g_a,h}}{1+\mu} 
\right) \left( \0{ \bar g}{\bar g_{a}  }
     \right)^{\012} \Biggr] \,. \notag\\
 \dot { \bar g}_{c}^{\ } 
  = &\,\left( 2+2 \eta_c +\bar\eta_h \right) \bar g_{c}^{\ } -
     \bar g_{c}^{2}\,\Biggl[ 
     c_{\bar g_c,c}+\0{d_{\bar g_c,c}}{1+\mu}  \notag \\
   &\,\hspace{.5cm}+\left( c_{\bar g_c,h}+ \0{d_{\bar g_c,h}}{1+\mu}\right) \left( \0{ \bar g}{\bar g_{c}  }
     \right)^{\012} \Biggr] \,. 
\label{eq:fullscalec}
\end{align} 
\end{subequations}
Here, the $d$ terms originate from the diagram with a regularised
graviton line, $(G\partial_t R_k G)^{(hh)}$.
The coefficients $c_{i,h}$ and $d_{i,h}$ are
$\lambda_3$ dependent as they receive contributions from the diagram with
a three-graviton vertex. The signs are chosen such that
$c_{i,h}(0) ,d_{i,h}(0)>0$.
The coefficients and the signs in the flow equation for $\bar g_c$ were not derived in this work.

The rescaled graviton anomalous dimension $\bar \eta_h$ reads
\begin{align}\label{eq:bareta}
  \bar\eta_h = - \0{\partial_t [Z_h (1+\mu)] }{Z_h (1+\mu) }= 
  \eta_h -\0{\dot \mu }{1+\mu}\,, 
\end{align} 
which includes the scale dependence of the full dressing of the
graviton propagator including the mass parameter. 
The set of anomalous dimensions is given by
\begin{align}
  \eta_h &= \bar g \left[c_{\eta_h,h}+
  \0{ d_{\eta_h,h}}{1+\mu} + (N_c^2-1) c_{\eta_h,a}
  \0{\bar g_a}{\bar g}\right]\,,\\
  \eta_c &=-\bar g  \left[c_{\eta_c,h} + \0{d_{\eta_c,h}}{1+\mu} \right] \,, \quad 
  \eta_a = - \bar g_a\left[ c_{\eta_a,h} - \0{d_{\eta_a,h}}{1+\mu} \right] \,, \notag
\end{align}
and completes the set of flow equations. Again, the graviton
contributions to $\eta_h$ have a $\lambda_3$ dependence with
$c_{\eta_h,h}(0), d_{\eta_h,h}(0)>0$. All other coefficients do not carry a 
$\lambda_3$ dependence.  
Note also that the $\partial_t \mu/(1+\mu)$ terms in the scaling terms
on the right-hand side of \eqref{eq:fullscale} come from the
normalisation of the $\bar g$'s with powers of $1/(1+\mu)$. In the
$\bar g_n$ flows this term is $n/(n-2)\partial_t \mu/(1+\mu)$ derived
from the rescaling \eqref{eq:bargng}. For the ghost-gravity and gauge
gravity couplings, it is always the term $\partial_t \mu/(1+\mu)$
derived from \eqref{eq:bargn}.

\section{Flow equations}
\label{app:floweq}
Here, we recall the results for the pure gravity flow for $\mu$,
$g_3$, and $\lambda_3$ derived in \cite{Christiansen:2015rva,Denz:2016qks},
add the derived gluon contributions,
and formulate them in terms of the rescaled couplings
\begin{align}
  \bar g_n &= g_n\left( \0{1}{1+\mu}\right)^{\0{n}{n-2}}\,,
  &
  \bar g_{c} &= g_{c} \left(\0{1}{1+\mu} \right)\,, \notag\\
  \bar g_{a} &= g_{a} \left(\0{1}{1+\mu} \right)\,, 
  &
  \bar \eta_h &= \eta_h -\0{\dot \mu}{1+\mu} \,,
\end{align}
see App.~\ref{app:scaling} and \eqref{eq:bargn} for details.
In order to show the interrelation of the different couplings we keep all
dependences on the higher couplings $\bar g_n$.
The flow equations are given by
\begin{widetext}
\begin{align} \nonumber 
\partial_t \mu  ={}& - \left( 2- \eta_h  \right) \mu
                     +\0{\bar g_3}{180 \pi} \Bigl[ 21 \left(10 - \eta_h\right) - 120 
                     \lambda_3 \left(8- \eta_h\right) +320 \lambda_3^2 \left(6- \eta_h\right) \Bigr] \\\nonumber 
                   & - \0{\bar g_4}{12 \pi} \Bigl[ 3 \left(8 - \eta_h\right) - 8 \lambda_4\left(6 - \eta_h\right) \Bigr]
                   - \left(1+\mu\right) \0{\bar g_c}{5 \pi} (10 - \eta_c) + \left(1+\mu \right) \left(N_c^2-1\right) \0{ \bar g_a \eta_a }{60 \pi}  \,,\\\nonumber
 \partial_t{\lambda}_3 ={}& - \left(1 + \0{\partial_t{\bar g}_3}{ 2 \bar g_3}-\032 \bar \eta_h\right) \lambda_3
		     + \bar g_3\Biggl\{-\01{1+\mu} \0{1}{240 \pi}\Bigl[ 11 \left(12 - \eta_h\right) 
                     - 72 \lambda_3 \left(10 -  \eta_h\right) + 120 {\lambda_3}^2 \left(8 - \eta_h\right) -
                     80\lambda_3^3 \left(6 - \eta_h\right) \Bigr] \\
                   & + \0{1}{6 \pi}\0{1}{1+\mu}\0{\bar g_4}{\bar g_3} \Bigl[3 \lambda_4 \left(8 - \eta_h\right) 
                     - 16 \lambda_3 \lambda_4 \left( 6 - \eta_h\right) \Bigr]
                     + \0{1}{8 \pi}\0{1}{1+\mu}\left(\0{\bar g_5}{\bar g_3}\right)^{\032}
                     \Bigl[ \left( 8 - \eta_h\right) - 4 \lambda_5 \left(6 -  \eta_h\right) \Bigr] 
                     \notag\\
                    & + \0{1}{10 \pi}\left(\0{\bar  g_c}{\bar g_3}\right)^{\032}  (12 - \eta_c)
                     + \0{1}{60 \pi}  \left(N_c^2-1\right)\left(\0{\bar  g_a}{\bar g_3}\right)^{\032}(3- \eta_a)\Biggr\}\,,\notag\\
\partial_t   \bar g_3 ={}& \left( 2 + 3 \bar \eta_h \right)  \bar g_3 
                      - \0{\bar g_3^2}{19 \pi}  \Bigg\{\0{1}{(1+\mu)^2}\0{2}{15} \Bigl[ 229- 1780 \lambda_3 + 3640 \lambda_3^2 - 2336 \lambda_3^3 \Bigr] \notag
                      \\ \nonumber
                   & - \0{1}{1+\mu}\0{1}{80} \Bigl[ 147 \left(10 - \eta_h\right) - 1860 \lambda_3 \left(8 - \eta_h\right) 
                     + 3380 \lambda_3^2 \left(6 - \eta_h\right) + 25920 \lambda_3^3 \left(4 - \eta_h\right) \Bigr]  \\\nonumber 
                 & -  \0{1}{1+\mu}\0{\bar g_4}{\bar g_3}\Biggl[\0{1}{18}\Bigl[ 45 \left(8 - \eta_h\right) - 8 \left( 30 \lambda_3  
                   - 59 \lambda_4 \right) \left(6 - \eta_h\right)- 360 \lambda_3\lambda_4 \left(4 - \eta_h\right) \Bigr] 
                   +\0{16}{1+\mu}\left( 1 - 3 \lambda_3 \right) \lambda_4 \Biggr] \\\nonumber 
                   & +\0{1}{1+\mu}\0{47}{6}\left(\0{\bar g_5}{\bar g_3}\right)^{\032} \left(6 - \eta_h\right) 
                     + \left(\0{\bar g_c}{\bar g_3}\right)^{\032} \Bigl[ \0{50 - 53 \eta_c}{10} \Bigr]
                     + \left(N_c^2-1\right)  \left(\0{\bar g_a}{\bar g_3}\right)^{\032} \Bigl[ \0{133 + \eta_a}{30}\Bigr]
                     \Biggr\}\,, \\\nonumber 
 \partial_t \bar g_a ={}& \left( 2 + 2 \eta_a + \bar \eta_h \right)  \bar g_a
		      - \0{\bar g_a^2}{30\pi} \Bigg\{ 
		      - \frac{100-13 \eta_a}{2}
		      + \frac{13 (5-\eta_h)}{\mu+1}
		      \notag\\
		      &+ \left(\0{\bar g_3}{\bar g_a}\right)^{\012}
		      \left(\frac{330-640 \lambda_3-\eta_a \left(33-80 \lambda_3\right)}{12}
		      + \frac{-15+400 \lambda_3-\eta_h \left(80 \lambda_3-6\right)}{3 (\mu+1)}\right)
		      \Biggr\}\,, 		      
 \label{eq:ggrav} 
\end{align}
and the anomalous dimension read 
\begin{align}
 \label{eq:anom-dim-ana}
 \eta_h&=\0{\bar g_3}{4\pi}\Bigg(\0{\bar g_4}{\bar g_3}(6-\eta_h)-\0{6 (8-\eta_h)+8 (6-\eta_h) \lambda_3-36 (4-\eta_h) \lambda_3^2}{9}
	      +\0{17+8 \lambda_3 (9 \lambda_3-8)}{3 (1+\mu)}
	      -\0{\bar g_c}{\bar g_3}\eta_c +\left(N_c^2-1\right) \0{\bar g_a}{\bar g_3}\0{1+\eta_a}{3}\Bigg)\,, \notag\\
 \eta_c &= -\frac{\bar g_c}{9\pi}\left(\frac{8-\eta_h}{1+\mu}+8-\eta_c\right) \,,
 \hspace{3cm}
 \eta_a =  -\0{\bar g_a}{8 \pi} \left(8-\eta_a - \0{4-\eta_h}{1+\mu} \right) \,.
\end{align}
\end{widetext}
The two terms in the flow equation for $\bar g_3$ proportional to
$1/(1+\mu)^2$ and the term in $\eta_h$ proportional to $1/(1+\mu)$
signal the derivative expansion at $p^2=0$. This is the
price to pay for an analytic flow equation.
On the other hand the terms proportional to $1/(1+\mu)$ in $\bar g_a$, $\eta_a$ and $\eta_c$ 
come from a regulator insertion in a graviton propagator 
compared to a ghost or gluon propagator.

The computation of these flow equations
involves contractions of very large tensor structures.
These contractions are computed with the help of the symbolic
manipulation system {\small \emph{FORM}}~\cite{Vermaseren:2000nd,Kuipers:2012rf}. 
We furthermore employ specialised Mathematica packages.
In particular, we use \emph{xPert}~\cite{xPert} for the generation of vertex functions,
and the \emph{FormTracer}~\cite{Cyrol:2016zqb} to trace diagrams.

\section{Coefficients in the scaling equations}
\label{app:coeffs}
The coefficients in the scaling equations in App.~\ref{app:scaling}
are given here in the
approximation \eqref{eq:bargid}. We assume that the anomalous
dimensions satisfy $|\eta|\leq 2$: they should not dominate the
scaling of the regulator. While the upper bound $\eta\leq 2$ is a
(weak) consistency bound for the regulator, for a detailed discussion,
see \cite{Meibohm:2015twa}, the lower one can be seen as a (weak)
consistency bound on the propagators. For $\eta<-2$, they cease to be
well-defined as Fourier transforms of space-time correlations
functions (if they scale universally down to vanishing momenta).
For simplicity, we display the coefficients with $\lambda_3=0$.
Note that all coefficients are defined such that they are always positive.
All coefficients can be directly read off from the equations \eqref{eq:ggrav} 
and \eqref{eq:anom-dim-ana}.

We get the coefficients $c_{\mu,h}$ and $c_{\mu,a}$ in the fixed
point equation of the mass parameter $\mu$ are given by
\begin{align}
  c_{\mu,h} &= \0{17}{6 \pi}-\0{2}{15 \pi} \eta_h-\0{1}{5 \pi}\eta_c \,,
  &
  c_{\mu,a} &= -\0{1}{60\pi} \eta_a\,.
\label{eq:cmu}
\end{align}
Note that the second coefficient is positive since $\eta_a<0$.
The coefficients $c_{\bar g,h}$ and $c_{\bar g,a}$ in the fixed point equation 
of the pure gravity coupling $\bar g$ read
\begin{align}
c_{\bar g,h} &= \0{47}{57 \pi}-\0{53}{190 \pi} \eta_h-\0{37 }{190 \pi}\eta_c\,, 
&
d_{\bar g,h} &= \0{598}{285\pi}\,,\notag \\
c_{\bar g,a} &=  \07{30\pi} +  \01{570\pi} \eta_a \,,
\label{eq:cg}
\end{align} 
while the coefficients $c_{\lambda_3,h}$ and
$c_{\lambda_3,a}$ in the fixed point equation of the 
coupling $\lambda_3$ are given by
\begin{align} 
c_{\lambda_3,h} &=  \0{33}{20 \pi} - \0{19}{240 \pi} \eta_h - \01{10 \pi} \eta_c\,, 
\notag\\
 c_{\lambda_3,a} &=  \0{3}{ 60 \pi} - \0{1}{ 60 \pi} \eta_a \,.  
\label{eq:cl}
\end{align}
Furthermore, the coefficient $c_{\bar g_a}$ in the fixed point equation for the 
two-gluon--graviton coupling $\bar g_a$ reads
\begin{align}
c_{\bar g_a,a} &=  \frac{5}{3 \pi }-\frac{13 }{60 \pi }\eta_a\,, 
&
d_{\bar g_a,a} &= \frac{13}{6 \pi } -\frac{13 }{30 \pi }\eta_h\,,
\notag\\
c_{\bar g_a,h} &=  \frac{11}{12 \pi } - \frac{11 }{120 \pi }\eta_a\,, 
&
d_{\bar g_a,h} &=  \frac{1}{6 \pi }-\frac{1}{15 \pi }\eta_h \,.
\end{align}
We also summarise the coefficients of the anomalous dimensions, to wit
\begin{align}\label{eq:of:canomalous} 
c_{\eta_h,h}&=  \0{1}{6 \pi} -\0{1}{12 \pi}\eta_h -\0{1}{4 \pi}\eta_c\,,
&
d_{\eta_h,h}&=  \0{17}{12\pi}\,,
\notag\\
c_{\eta_h,a}&=  \0{1}{12\pi}+\0{1}{12\pi}\eta_a\,,
&&
\\
c_{\eta_c}&=  \0{8}{9\pi} - \0{1}{9\pi}\eta_c \,, 
&
d_{\eta_c}&=  \0{8}{9\pi}  -\0{1}{9\pi}\eta_h \,, 
\notag\\
c_{\eta_a}&=  \01{\pi} - \01{8\pi}\eta_a   \,,
&
d_{\eta_a}&=  \01{2\pi} - \01{8\pi} \eta_h\,.
\notag
\end{align}

\bibliography{flatgravity}
\end{document}